\documentclass[fleqn,usenatbib]{mnras}

\usepackage{newtxtext,newtxmath}
\usepackage[T1]{fontenc}

\DeclareRobustCommand{\VAN}[3]{#2}
\let\VANthebibliography\thebibliography
\def\thebibliography{\DeclareRobustCommand{\VAN}[3]{##3}\VANthebibliography}

\usepackage{multirow}
\usepackage{booktabs}
\usepackage{pdflscape}
\usepackage{graphicx}	% Including figure files
\usepackage{amsmath}	% Advanced maths commands
\usepackage{ulem}
\title[Sco X-1 as a continuous waves source]{Sco X-1 as a continuous gravitational waves source: modelling the secular evolution using MESA}

\author[G. Pagliaro et al.]{
Gianluca Pagliaro,$^{1,2}$\thanks{E-mail: gianluca.pagliaro@aei.mpg.de}
Maria Alessandra Papa,$^{1,2}$\thanks{E-mail: maria.alessandra.papa@aei.mpg.de}
Jing Ming,$^{1,2}$
and Devina Misra$^{3}$
\\
$^{1}$Max Planck Institute for Gravitational Physics (Albert Einstein Institute), Callinstr. 38, 30167 Hannover, Germany\\
$^{2}$Leibniz Universit\"at Hannover, D-30167 Hannover, Germany\\
$^{3}$Department of Physics, Norwegian University of Science and Technology, NO-7491 Trondheim, Norway
}

\date{Accepted XXX. Received YYY; in original form ZZZ}

\pubyear{\the\year{}}

\begin{document}
\label{firstpage}
\pagerange{\pageref{firstpage}--\pageref{lastpage}}
\maketitle

\begin{abstract}
We study the prospects for detecting continuous gravitational waves from Sco X-1, evaluating the most likely waveform and progenitor parameters. We model the spin of the neutron star by the accretion torque and the gravitational-wave torque, considering two mechanisms for generating the non-axisymmetry responsible for the latter: magnetic mountains and crustal breakage deformation. Both torques are intertwined with the binary evolution, which we trace from the formation of the neutron star in a binary system with a main-sequence companion. We do this with MESA, starting from a set of initial binary configurations.
At LIGO-O3 sensitivity, a magnetic ellipticity of $\varepsilon \gtrsim 10^{-6}$ is necessary for detection.
The highest frequency at which we have detectable signals increases with the accretion efficiency $\eta$; it is as high as 360 Hz.
At 3G (Cosmic Explorer/Einstein Telescope) sensitivity, ellipticities as small as $6 \times 10^{-9}$, are detectable, but the waveform highly depends on the binary system: the highest frequency of detectable signals spans the very broad range 600-1700 Hz, strongly depending on $\eta$ and mass of the progenitor donor star $M^d$.
In Sco X-1-like systems with $\eta \leq 30$ per cent, the crust does not break. For $\eta \in$[40 \textrm{per cent}, 60 \textrm{per cent}], only progenitors with $M^d \geq$[1.1, 1.5]$M_{\odot}$ present crustal breakage. In some systems, the crust breaks during their Sco X-1-like phase. If Sco X-1 were one of those systems, it would emit strong gravitational waves sweeping from $\mathcal{O}$(1000)Hz down to torque-balance frequencies in $\approx 150\,000 \, [ \varepsilon / 10^{-5} ]^{-2/5}$ yr. 
We estimate the current detection probability for this signal to be well below 1 per cent; this probability increases substantially -- to around 41 per cent -- with 3G detectors.
\end{abstract}


%%%%%%%%%%%%%%%%%%%%%%%%%%%%%%%%%%%%%%%%%%%%%%%%%%

%%%%%%%%%%%%%%%%% BODY OF PAPER %%%%%%%%%%%%%%%%%%

\section{Introduction}

Scorpius X-1 (also Sco X-1) is a persistent low mass X-ray binary consisting of a neutron star accreting matter from a roche-lobe filling donor.
The release of gravitational energy possessed by the infalling matter makes the system visible in the X-spectrum.
The matter, spiralling-in from an accretion disc on to the surface of the neutron star, transfers angular momentum to it, spinning it up.
This picture is not unique to Sco X-1, and virtually applies to all low mass X-ray binaries hosting a neutron star and is referred to as the recycling scenario \citep{1982CSci...51.1096R, 1982Natur.300..728A, 1991PhR...203....1B}. Neutron stars that have completed this process, now visible as rapidly rotating pulsars, are known as recycled pulsars. Sco X-1, and in particular the neutron star it hosts, does not fall within the latter category, as no pulsations have been so far detected \citep{2022MNRAS.509.1745G, 2025arXiv250804873L}.

Even before the consolidation of this theoretical scheme with the discovery of the accreting millisecond X-ray pulsar SAX J1808.4-3658 \citep{1998Natur.394..344W}, the lack of pulsars spinning above $\approx \! 700 \, \textrm{Hz}$ started to pose an astrophysical puzzle hard to reconcile with standard mass accretion models, which predicted recycling frequencies above this observational cut-off \citep{1988ApJ...324..363W}.

\citet{1998ApJ...501L..89B} conjectured that it is the emission of continuous gravitational waves from the accreting neutron star that provides the counterbalancing spin-down torque preventing these systems to reach spins higher than those observed. They continue speculating that Sco X-1 will likely be the first to ever be observed emitting such type of gravitational waves. 

The reasoning behind Bildsten's conjecture has to do with the fact that the spin-down torque exerted by the emission of continuous gravitational waves grows with the fifth power of the star's angular velocity, limiting de facto the recycling frequency. 
Moreover, the non-axisymmetric mass distribution on the neutron star, necessary for gravitational wave emission to occur, may have its cause in the intricate process of mass accretion.
Within this context, rapidly accreting systems should be the brightest gravitational wave sources, because the torque to counterbalance is correspondingly higher (details in Section \ref{sec:torques}). To date, Sco X-1 is the most rapidly accreting persistent low mass X-ray binary known.

Searches for continuous gravitational waves from Sco X-1 have been carried out almost continuously for about two decades. 
From the first search of a periodic signal in LIGO S2 data \citep{2007PhRvD..76h2001A} to the most recent ones in LIGO's third observational run \citep{2022ApJ...941L..30A, 2022PhRvD.106f2002A, 2023ApJ...949..117W, 2025PhRvD.111h4040V}, searches have greatly improved in sensitivity and frequency coverage.
Given that a detection is still missing, search results consist primarily in upper limits on the wave amplitude throughout the frequency band surveyed.

In addition to observational results, a refinement in accretion models have to some extent scaled down the prospects to detect continuous gravitational waves from accreting neutron stars, and by implication from Sco X-1.
\citet{2005MNRAS.361.1153A} and \citet{2012ApJ...746....9P} for example, show that properly accounting for disc-magnetosphere interaction could be enough to reconcile with the observational spin frequency cut-off of $\approx 700 \, \textrm{Hz}$, without having to rely on gravitational wave emission. 
On the same issue, \citet{2015MNRAS.451.2117B} speculate it is the neutron star's magnetic field that, as a result of its decay, ceases to funnel the accretion flow on to the neutron star and the system becomes detached before reaching sub-millisecond spin periods.

This, however, does not mean accreting neutron stars do not radiate continuous gravitational waves. 
Several viable deformation mechanisms are indeed plausibly at work in such systems \citep{1998ApJ...501L..89B, 1999ApJ...516..307A, 2000MNRAS.319..902U, 2005ApJ...623.1044M, 2024PhRvD.110d4016M, 2025ApJ...978L...8M}.
Moreover, due to its very high mass accretion rate, rather than a typical low mass X-ray binary, Sco X-1 is an outlier difficult to fit into the analysis of \citet{2012ApJ...746....9P} for instance, as the authors themselves recognise.
Precisely for these reasons, Sco X-1 remains one of the most interesting potential sources of continuous gravitational waves worth to study.

It is useful in our opinion to investigate on both the influence that the long-term evolution of Sco X-1 has on continuous gravitational wave emission, and, as a byproduct, on aspects related to gravitational wave searches such as parameter space coverage and detection probability, taking into account different emission scenarios.

An issue which has received little to no attention so far is the effect that binary parameters, such as mass of the donor star or the efficiency of the mass transfer, have on the resulting gravitational wave emission. Recent studies on the long-term evolution of Sco X-1 find indeed it to be compatible with a moderately wide range of binary configurations at birth, which turn out to have markedly different evolutionary pathways. Whether, and if so how, this influences the emission of gravitational waves is unclear.

To investigate on this we simulate the evolution of Sco X-1 in all its parts using MESA \citep{2011ApJS..192....3P, 2013ApJS..208....4P, 2015ApJS..220...15P, 2018ApJS..234...34P, 2019ApJS..243...10P, 2023ApJS..265...15J}, extending the framework introduced in \citet{2025AA...693A.314M} to account for gravitational wave emission from the neutron star.
With respect to the latter, we model two different deformation mechanisms, carefully studying the output of our simulations for each of them separately.
In doing so, we identify the main effects that different binary configurations have on quantities that play a role in the emission and detection of continuous gravitational waves.
We verify the adequacy of currently employed parameter space coverage in gravitational wave searches, also estimating the probability to detect gravitational waves from Sco X-1, accounting for current and future sensitivity.

The paper is structured as follows: in Section \ref{sec:known_params} we state all the relevant quantities pertaining to Sco X-1 which are subject to observational constraints. In Sections \ref{sec:bin_evol_mesa} and \ref{sec:ns_spinevol} we describe the implementation of the binary evolution and the spin frequency evolution of the accreting neutron star respectively. In Section \ref{sec:def_mech} we detail the two deformation mechanisms modelled. Our main results are presented and discussed in Section \ref{sec:results}, while we draw our conclusions in Section \ref{sec:discussion}.

\section{Observational constraints on Sco X-1}
\label{sec:known_params}

Since its discovery, Sco X-1 has held the record as the brightest extra solar X-ray source in the sky \citep{1962PhRvL...9..439G}.
Parallax measurements place it at a distance of $2.8 \pm 0.3 \, \textrm{kpc}$ and its X-ray luminosity is estimated to be $L_X = 2.3 \times 10^{38} \textrm{erg} \textrm{s}^{-1}$ \citep{1999ApJ...512L.121B}.
The orbital period is known with very good accuracy being $P = 0.7873139 \pm 0.0000007 \, \textrm{d}$ \citep{2023MNRAS.520.5317K}.
The mass ratio between donor and neutron star lays within the range $0.28 < q < 0.51$ \citep{2015MNRAS.449L...1M}.
The effective temperature of the donor star has an upper limit of $T_{\textrm{eff}} < 4800 \, \textrm{K}$ \citep{2015MNRAS.449L...1M, 2016MNRAS.456..263P}.
Extensive radio observations estimate the orbital inclination angle to be $\iota = 44^{\circ} \pm 6^{\circ}$ \citep{2001ApJ...558..283F}.

The X-ray luminosity can be converted into a quantity of greater practical use in the context of this paper, the mass accretion rate. 
\citet{2008MNRAS.389..839W} estimates the latter via the bolometric flux, obtaining
\begin{equation}
\label{eq:mdot_scox1_bol}
\dot{M}_{\textrm{a}}^{\textrm{bol}} = 3.1 \times 10^{-8} \left( \frac{R}{10 \, \textrm{km}} \right) \, \left( \frac{1.4 \, M_{\odot}}{M} \right) M_{\odot} \, yr^{-1}
\end{equation}
where $R$ and $M$ are respectively radius and mass of the neutron star. In Equation \ref{eq:mdot_scox1_bol}, and for the rest of the paper, the dot on top of any quantity indicates its time derivative.

Sco X-1 is also visible in the UV, primarily due to re-emission from the inner disc as it is being heated up by X-rays. \citet{1990AA...235..162V, 1991ApJ...376..278V} model this process and infer the mass accretion rate consistent with UV data. The average mass accretion rate between three different accretion states reads
\begin{equation}
\label{eq:mdot_scox1_uv}
\dot{M}_{\textrm{a}}^{\textrm{UV}} = 1.6 \times 10^{-8} \left( \frac{R}{10 \, \textrm{km}} \right) \, \left( \frac{1.4 \, M_{\odot}}{M} \right) M_{\odot} \, yr^{-1} .
\end{equation}

We define a {\it{Sco X-1 resemblance}} region as the following 4-dimensional cuboid
\begin{equation}
\label{eq:cuboid}
\mathcal{C} = 
	\begin{cases}
	0.7586 \leq P/[\textrm{d}] \leq 0.8511 \\
	0.28 \leq q \leq 0.51 \\
	1.38 \times 10^{-8} \leq \dot{M}_{\textrm{a}}/[M_{\odot} \textrm{yr}^{-1}] \leq 3.87 \times 10^{-8} \\
	T_{\textrm{eff}} < 4800 \, \textrm{K} ,
	\end{cases}
\end{equation}
and consider in what follows a simulated binary to resemble Sco X-1, if at any time of its evolution it enters the hypercuboid defined as above.

The range in mass accretion rate brackets the two values of Equations \ref{eq:mdot_scox1_bol} and \ref{eq:mdot_scox1_uv}, accounting for different mass-radius combinations (see second paragraph of Section \ref{sec:ns_spinevol}).
We choose a rather loose interval for Sco X-1 compatible orbital periods, the same as considered by \citet{2021ApJ...922..174V} (see Table 2 of their work).
This choice was necessary in order to maintain simulation times at an acceptable level.

\section{Binary evolution with MESA}
\label{sec:bin_evol_mesa}
We simulate the evolution of Sco X-1 using MESA in its 24.08.1 release version.
MESA is an open-source 1D stellar evolution code mainly employed to perform detailed evolution of internal and global properties of isolated and binary stars, as well as to compute the dynamics of binary stars experiencing mass transfer.
Our aim here is to evolve Sco X-1 as a binary evolving into a semi-detached system leading to mass accretion on to the neutron star, which in turn changes its spin state and develops a net non-axisymmetry.

\subsection{The MESA grid}
\label{sec:mesa_grid}
Since we know Sco X-1 only in its present state, we adopt the reverse population synthesis technique introduced by \citet{2021ApJ...922..174V}. This consists of setting up a grid of free parameters that characterise Sco X-1 at birth, where by birth we mean the birth of the neutron star, with the companion star in its zero age main sequence. We keep track of observationally constrained quantities such as orbital period, mass ratio, mass accretion rate and temperature of the donor star, flagging any system as a Sco X-1 {\it{progenitor}} (referring to its properties at the start of the simulation) if at any moment of its evolution it resembles Sco X-1.

The free parameters characterising the binary at birth are orbital period $P$ and mass of the donor star $M^{\textrm{d}}$.
We profit from the results of \citet{2021ApJ...922..174V}, and consider a relatively small $P-M^{\textrm{d}}$ space. The grid in the orbital period dimension is uniform in logarithm, comprised of 36 bins within $-0.1 \leq \log_{10} \, P/[\textrm{d}] \leq 0.6$, while that in donor masses is uniform within $1.0 \leq M^{\textrm{d}}/M_{\odot} \leq 2.0 $ with a step of $0.025 M_{\odot}$. The 2-dimensional $P-M^{\textrm{d}}$ space is therefore comprised of a total of 1476 bins.

An additional free parameter we account for is the accretion efficiency, which is the fraction of mass lost by the donor that is actually accreted by the neutron star. In general, the accretion efficiency is never equal to one, and some mass is inevitably lost by the system, with a wide range of efficiencies plausibly realised in low mass X-ray binaries \citep{1999AA...350..928T, 2025AA...693A.314M, 2024MNRAS.535..344K}.
In these systems, the most important mechanism to account for in this regard is fast wind from the vicinity of the accretor \citep{2023pbse.book.....T}.

As the neutron star spins-up, the star can indeed enter a "propeller phase", and partly or entirely eject matter from the boundary between the magnetosphere and the inner disc \citep{1975AA....39..185I}.
An exhaustive description of the phenomenon is difficult to obtain, although the underlying physics is broadly understood and can be traced back to the seminal works of \citet{1977ApJ...217..578G, 1979ApJ...232..259G, 1979ApJ...234..296G}.
In addition, there are concurrent mechanisms such as radiation driven outflows \citep{1986ApJ...302..519P}, disc winds \citep{1982MNRAS.199..883B}, disc instabilities \citep{2010MNRAS.406.1208D} that further complicate an appropriate modelling of non-conservative mass transfer in low mass X-ray binaries.

We do not attempt to model the propeller phase nor any of the mechanisms listed above, rather we account for simulating Sco X-1 at various average accretion efficiencies. 
This is done in MESA by specifying a value for $\beta$, a parameter corresponding to the percentage of the infalling matter that is ejected by the accretor as isotropic fast wind. The accretion efficiency reads therefore
\begin{equation}
\label{eq:acc_eff}
\eta = 1 - \beta .
\end{equation}
We set a grid in $\beta$ from $0.8$ to $0.2$ with a step of $-0.1$, correspondent to accretion efficiencies between $20\%$ and $80\%$ and a discrete step of $10\%$. The 3-dimensional $P-M^{\textrm{d}}-\beta$ (or equivalently $P-M^{\textrm{d}}-\eta$) grid therefore consists of 10332 bins.

Different values of the accretion efficiency affect both the orbital evolution (due to different amount of mass loss), but also trivially the rate at which mass is accreted. We have indeed
\begin{equation}
\label{eq:macc_mtransf}
\dot{M}_{\textrm{a}} = \eta \, \dot{M}_{\textrm{tr}}
\end{equation}
with $\dot{M}_{\textrm{tr}}$ being the mass transfer rate, namely the rate at which mass is transferred from the donor star to the accretion disc. 
In Equation \ref{eq:macc_mtransf} one must be aware that the quantity on the left is the baryonic mass accretion rate. Part of the rest mass of the accreted matter will indeed be converted into radiation, and the gravitational mass accreted by the neutron star will be a fraction of the baryonic one. We account for this effect assuming that $87.5\%$ of the baryonic mass will be retained by the neutron star as gravitational mass \citep{2007PhR...442..109L}.

In this regard we acknowledge the work of \citet{2024MNRAS.535..344K, 2025ApJ...980...51K} who account for a more realistic dynamic baryonic-to-gravitational mass conversion.

\subsection{Orbital angular momentum evolution}
The dynamical evolution of the binary system is governed by the following equation
\begin{equation}
\label{eq:jdot}
\dot{J} = \dot{J}_{gr} + \dot{J}_{mb} + \dot{J}_{ml} 
\end{equation}
where $J$ is the total angular momentum of the binary system. 

$\dot{J}_{gr}$ is the orbital angular momentum rate of change due to gravitational wave radiation. We warn against confusing it with the gravitational wave radiation emitted by the neutron star.
We use the default MESA treatment for this quantity (see \citet{2015ApJS..220...15P}).

Also for $\dot{J}_{ml}$, the orbital angular momentum change due to mass loss, we use MESA default prescription \citep{2015ApJS..220...15P} accounting for mass ejected in the vicinity of the accretor only, controlling this effect through $\beta$ as discussed in the previous section. 
In the general case we have indeed
\begin{equation}
\label{eq:acc_eff_general}
\eta = 1 - \alpha - \beta - \delta
\end{equation}
where $\alpha$ and $\delta$ are the fractions of mass ejected respectively in the vicinity of the donor (direct fast wind) and a circumbinary coplanar toroid. We set both parameters to zero adopting the isotropic re-emission prescription that assumes a conservative matter flow from the donor to the accretor \citep{1991PhR...203....1B}.

Mass will be lost by the system also if the accretion rate exceeds the Eddington limit. Following \citet{2023pbse.book.....T}, we calculate the latter to be
\begin{equation}
\label{eq:mdot_edd}
\dot{M}_{\textrm{Edd}} = \frac{4 \pi R c}{0.2 (1 + X)}
\end{equation}
where $X$ is the hydrogen mass fraction of the accreted matter, calculated by MESA at each timestep.

Depending on mass and radius of the neutron star, and on $X$, the quantity expressed by Equation \ref{eq:mdot_edd} can be bigger or smaller than the observational limit on mass accretion rate of Equation \ref{eq:mdot_scox1_bol}. We limit the rate of accretion to $\dot{M}_a \leq max\{\dot{M}_{\textrm{Edd}}, \dot{M}_a^{bol}\}$, therefore allowing Sco X-1 to accrete at "mild" super Eddington accretion rates. 

$\dot{J}_{mb}$ is the angular momentum rate of change due to magnetic braking. The mechanism behind magnetic braking has to do with the fact that low mass stars (in the case here considered, the donor) with convective envelopes produce winds forced by the stellar magnetic field to co-rotate with the star out to a certain distance. This provides a significant drain of spin angular momentum of the donor star. In a close low mass binary however, tidal forces are strong enough to maintain the donor star in spin-orbit coupling, therefore as the donor star loses spin angular momentum, tidal forces spin it back up. This is done at the expense of the orbital angular momentum, and the orbit shrinks.

We use the CARB magnetic braking prescription which is, in contrast to the default MESA implementation (Skumanich law, see \citep{1983ApJ...275..713R}), able to faithfully reproduce rapidly accreting systems like Sco X-1 \citep{2019MNRAS.483.5595V}.
Since the CARB magnetic braking law is expressed by a rather complex equation, we refer the reader to the original article \citep{2019ApJ...886L..31V}.

\section{Evolution of the neutron star}
\label{sec:ns_spinevol}
We consider the neutron star in Sco X-1 to be born as a "typical" pulsar with mass $M_0 = 1.4 \, M_{\odot}$, dipolar magnetic field $B_0 = 10^{12} \, \textrm{G}$, and birth spin frequency $\nu_0 = 300 \, \textrm{Hz}$. Since we assume that accretion is the cause of the quadrupolar deformation sourcing gravitational wave emission, up until the start of mass transfer, our neutron star spins down only due to magneto-dipolar braking.
Its angular velocity at the onset of mass transfer is
\begin{equation}
\label{eq:magneto_dipolar_analytic}
\Omega = \left( \frac{1}{\Omega_0^2} + \frac{4}{3} \frac{R^6 B^2}{c^3 I} \Delta t \right)^{-1/2}
\end{equation}
where $\Omega = 2 \pi \nu$ as usual, and $\Delta t$ is the time elapsed from the birth of the neutron star to the start of mass transfer. Equation \ref{eq:magneto_dipolar_analytic} -- valid in cgs units -- is easily obtained by analytic integration of the standard magneto-dipolar spin-down torque (see, for example, \citet{Lyne_Graham-Smith_2012}).

As the mass of the neutron star changes due to accretion, the radius is computed according to the AP3 neutron star equation of state \citep{1998PhRvC..58.1804A}. We extract tabulated data from LALsimulation, part of LALSuite software package \citep{lalsuite}. We choose this particular EOS because of a recent model selection study which singles out this one to best accomodate observational results on tidal deformability from neutron star coalescence gravitational wave events, as well as NICER observations on a set of pulsars \citep{2022ApJ...926...75B}. The radius at $1.4 \, M_{\odot}$ for AP3 EOS is $\approx 12 \, \textrm{km}$.

Recently, \citet{2023MNRAS.524.4899C} have proved the relevance of the variation of the moment of inertia in the spin evolution of accreting neutron stars.
As mass and radius change then, we compute the moment of inertia of the neutron star using the following Equation
\begin{equation}
\label{eq:mom_iner}
I = 0.237 \times M R^2 \left[ 1 + 4.2 \frac{M}{M_{\odot}} \frac{\textrm{km}}{R} + 90 \left( \frac{M}{M_{\odot}} \frac{\textrm{km}}{R} \right)^4 \right] 
\end{equation}
taken from \citet{2005ApJ...629..979L}, who show it to adequately approximate the moment of inertia of neutron stars with $M > 1.0 \, M_{\odot}$ and a non-excessive degree of softening, i.e. EOSs with an upper bound on mass significantly above $1.6 \, M_{\odot}$ (the upper bound on mass for the AP3 EOS is of about $2.4 \, M_{\odot}$).

Accreted mass also buries magnetic field lines deep within the neutron star, weakening the dipolar field. We use the phenomenological relation
\begin{equation}
\label{eq:mag_field_decay}
B = \frac{B_0}{1 + \Delta M / m_B}
\end{equation}
obtained by \citet{1989Natur.342..656S}, where $\Delta M$ is the accreted mass, while $m_B$ is a parameter that, at fixed mass accretion rate, controls how quickly the dipolar field decays. We fix this quantity to $m_B = 10^{-4} \, M_{\odot}$, which is the value the authors find to consistently reproduce the evolutionary tracks of known millisecond radio and X-ray sources. 

\subsection{Torques at work on the accreting neutron star}
\label{sec:torques}
Conservation of spin angular momentum implies
\begin{equation}
\label{eq:ang_mom_cons}
\dot{\Omega} = \frac{\tau_{tot}}{I} - \frac{\dot{I}}{I} \Omega
\end{equation}
where $\tau_{tot}$ is the total torque acting on the neutron star, which we conveniently split as follows
\begin{equation}
\label{eq:tot_torque_components}
\tau_{tot} = \tau_{acc} + \tau_{cgw} .
\end{equation}

The first term on the right hand side of Equation \ref{eq:tot_torque_components} is the accretion torque exerted by the infalling matter, accounting also for its interaction with the neutron star's magnetosphere.
Since spin-up happens due to angular momentum exchange between the accretion disc and the neutron star, we assume $\tau_{acc}$ to act on the neutron star only if
\begin{equation}
\label{eq:dubuis}
\begin{split}
\dot{M}^{tr} \geq \dot{M}^{crit} := \, & 3.2 \times 10^{-9} \left( \frac{M}{1.4 \, M_{\odot}} \right)^{1/2} \\ 
&\left( \frac{M^d}{1 \, M_{\odot}} \right)^{-1/5} \left( \frac{P}{1 \, \textrm{d}} \right)^{7/5} \, M_{\odot} \textrm{yr}^{-1}
\end{split}
\end{equation}
which ensures the stability of the accretion disc \citep{1999MNRAS.303..139D}.
Notice that we allow the neutron star to accrete mass at the specified average accretion efficiency even if $\dot{M}^{tr} < \dot{M}^{crit}$ (differently than what done for example by \citet{2025AA...693A.314M}).

\citet{2006AA...451..581D} modify the standard accretion model in order to account for the observationally motivated phenomenon of accretion during the propeller phase (see \citet{2006AA...451..581D} and references therein). Whether or not Sco X-1 is in the propeller phase, we know it is currently accreting mass, therefore this model is appropriate to describe it up to its current state.
Their accretion torque reads
\begin{equation}
\label{eq:acc_torque}
\tau_{acc} = 
	\begin{cases}
	\dot{M}_\textrm{a}(GMR_{i})^{1/2} \left[(1 - \omega) + \frac{\sqrt{2}}{3} \left( 1 - 2 \omega + \frac{2 \omega^2}{3} \right) \right], & if \, \omega \leq 1 \\
	\dot{M}_\textrm{a}(GMR_{i})^{1/2} \left[(1 - \omega) + \frac{\sqrt{2}}{3} \left(\frac{2}{3 \omega} - 1 \right) \right], & if \, \omega > 1 ,
	\end{cases}
\end{equation}
where $\omega$ is the {\it{fastness parameter}}, treated later in this section, and $R_i$ is the {\it{inner radius}}, which is the radius at which the accretion disc is truncated. 
In \citet{2006AA...451..581D} the latter is taken to be equal to the magnetospheric radius
\begin{equation}
\label{eq:mag_rad}
R_m = \left( \frac{ \left( B R^3 \right)^4 }{2 G M \dot{M}_\textrm{a}^2} \right)^{1/7} 
\end{equation}
which is where the pressures exerted radially by the magnetic dipole and the accreting plasma balance. 

Recently, \citet{2023MNRAS.524.4899C} have pointed out that in rapidly accreting massive neutron stars, the disc might be truncated at the neutron star's ISCO radius (a widely known fact for accreting black holes) if this happens to become bigger than both the neutron star and the magnetospheric radius. 
In practise, the ISCO radius of a $1.4 \, M_{\odot}$ neutron star lays comfortably within the star's surface, but as the neutron star grows in mass, the ISCO radius also grows, eventually "leaking" outside the neutron star, and, for rapidly accreting systems with a relatively weak magnetic dipole, it can become bigger than the magnetospheric radius. Under these conditions, in the region between the radius of the neutron star and the ISCO radius, the matter cannot co-rotate with the disc, so the disc is truncated at the ISCO.
\citet{2018ApJ...861..141L} numerically find a universal relation for the ISCO radius of a rotating neutron star 
\begin{equation}
\label{eq:isco_rad}
\begin{split}
R_{ISCO} &= 8.81 \left( \frac{M}{M_{\odot}} \right) - 1.46 \times 10^{-4} \left( \frac{M}{M_{\odot}} \right)^2 \Omega \\
	& + 2.22 \times 10^{-9} \left( \frac{M}{M_{\odot}} \right)^3 \Omega^2 - 2.43 \times 10^{-14} \left( \frac{M}{M_{\odot}} \right)^4 \Omega^3 \ \textrm{km} .
\end{split}
\end{equation}
For instance, at $M = 1.65 \, M_{\odot}$ and $\frac{\Omega}{2 \pi} = 500 \, \textrm{Hz}$, the above equation gives $R_{ISCO} \approx 13 \, \textrm{km}$, which is greater than the radius at any mass for the AP3 equation of state.
We therefore define
\begin{equation}
\label{eq:inner_radius}
R_i = max \{ R, R_m, R_{ISCO} \} .
\end{equation}

The accretion torque of Equation \ref{eq:acc_torque} contains the fastness parameter $\omega$, defined to be the ratio between the angular velocity of the star and the Keplerian angular velocity at the inner radius, namely
\begin{equation}
\label{eq:fastness_param}
\omega = \Omega \left( \frac{R_i^3}{G M} \right)^{1/2} .
\end{equation}
Understanding this quantity in the context of the standard accretion scenario is useful.
The Keplerian angular velocity at the inner radius is the upper limit to the star's angular velocity above which matter will simply {\it{escape}} the inspiralling orbit around the neutron star, and get ejected. Therefore, as the angular velocity increases and equals the Keplerian velocity at the inner radius ($\omega \approx 1$), the neutron star enters the propeller regime.
In \citet{2006AA...451..581D} this transition is not as sharp, matter can still be accreted in the propeller phase and the torque changes as Equation \ref{eq:acc_torque} shows (note also how the torque becomes negative before this phase, already when $\omega \approx 0.9$). 

Finally, $\tau_{cgw}$ is the continuous gravitational wave torque slowing down the neutron star. This is
\begin{equation}
\label{eq:cgw_torque}
\tau_{cgw} = - \frac{32 G I^2 \varepsilon^2}{5 c^5} \Omega^5 .
\end{equation}

The quantity $\varepsilon$, called equatorial ellipticity, is the degree of non-axisymmetry with respect to the spin axis, and is defined as
\begin{equation}
\label{eq:ellipticity}
\varepsilon = \frac{| I_{xx} - I_{yy} | }{I_{zz}}
\end{equation}
where $I_{ii}$ are the diagonal components of the inertia tensor of the neutron star, and $z$ is chosen to be the spin axis (the next section will be devoted to discuss the deformation mechanisms responsible for non-zero ellipticities in Sco X-1).
As a result, the neutron star will emit a continuous gravitational wave of amplitude
\begin{equation}
\label{eq:h0}
h_0 = \frac{4 \pi^2 G I}{c^4} \frac{ f^2 \varepsilon}{d} 
\end{equation}
at frequency $f = \Omega / \pi$, i.e. at twice the spin frequency of the neutron star. In the above, $d$ is the distance to the system.

A concept that will later come in handy is that of torque-balance.
As the neutron star in Sco X-1 is spun-up to high $\Omega$, the quantity in Equation \ref{eq:cgw_torque} rapidly grows, and, at some point, balances out the spin-up torque.
This is clearly a dynamically evolving condition, but it is often useful to consider it at present time, making a few simplifying assumptions.
If we approximate the accretion torque taking $\tau_{acc} \approx \dot{M}(GMR_{i})^{1/2}$, we can plug-in Equation \ref{eq:h0} into Equation \ref{eq:cgw_torque}, equate the accretion torque with the absolute value of the gravitational wave torque to obtain
\begin{equation}
\label{eq:torque_balance_general}
	\begin{split}
	h_0^{tb} = 1.49 \times 10^{-24} & \left( \frac{\dot{M}_\textrm{a}}{3.1 \times 10^{-8} M_{\odot} yr^{-1}} \right)^{1/2} \left( \frac{M}{1.4 M_{\odot}} \right)^{1/4} \\
	& \left( \frac{R_i}{10 \textrm{km}} \right)^{1/4} \left( \frac{\Omega}{\textrm{Hz}} \right)^{-1/2} . 
	\end{split}
\end{equation}

$h_0^{tb}$ is the amplitude of the gravitational wave emitted by the neutron star in Sco X-1 that has reached torque balance at the angular velocity $\Omega$, given the value of mass accretion rate at present time, assuming the accretion disc to be truncated at a distance $R_i$ from the centre of the neutron star.
In what follows, when quantities in brackets of Equation \ref{eq:torque_balance_general} are equal to 1, we refer to {\it{nominal torque balance}}.

\section{Deformation mechanisms}
\label{sec:def_mech}
We model two distinct deformation mechanisms at work on the accreting neutron star. They are described in detail below.

\subsection{Magnetic confinement of accreted matter}
\label{sec:mag_conf}
Initially, the accreted matter, funneled by the dipolar magnetic field lines, lands on the magnetic poles, rapidly sinking and redistributing itself, migrating towards the equatorial region of the neutron star. As matter spreads towards the equator, the magnetic field lines get screened diamagnetically, "diluting" the magnetic field at the poles, and amplifying it at the equator (this is exactly the burial of the dipolar magnetic field).
As the equatorial magnetic field lines become more and more compressed, the accreted matter moves towards the equator with greater difficulty, and an equilibrium configuration is reached with the neutron star developing "mountains" at the poles, resulting in a net non-axisymmetry if the magnetic field axis and spin axis are misaligned.
\citet{2005ApJ...623.1044M} model this process, giving an equation to calculate the resulting ellipticity that reads as follows
\begin{equation}
\label{eq:melatospayne}
\varepsilon = \frac{5}{2} \frac{\Delta M}{M} \left( 1 - \frac{3}{2 b} \right) \left( 1 + \frac{\Delta M b^2}{8 M_c} \right)^{-1} .
\end{equation}
In the above, $\Delta M$ is the amount of mass accreted and $M$ is the mass of the neutron star. 

The quantity $M_c$ is the "critical accreted mass". Its value depends on the natal magnetic field as well as on the temperature of the crust. 
\citet{2005ApJ...623.1044M} calculate that for a "typical" neutron star similar to the one defined at the beginning of Section \ref{sec:ns_spinevol}, $M_c \approx 10^{-4}$, which is the value we fix for this quantity.

Equation \ref{eq:melatospayne} describes a deformation that rapidly grows as matter is accreted, and very early on the accretion phase saturates for $\Delta M \gtrsim 8 M_c / b^2$. It does not apply in the large-$\Delta M$ regime ($\Delta M \gtrsim 0.1 \, M_{\odot}$) as the theoretical framework used to obtain it does not hold. 
We assume that, as the neutron star has accreted the saturation mass $\Delta M = 8 M_c / b^2$, it maintains a constant "saturation ellipticity" for the rest of the evolution. We compute the ellipticity as follows
\begin{equation}
\label{eq:melatospayne_adapted}
\varepsilon = 
	\begin{cases}
	\frac{5}{2} \frac{\Delta M}{M} \left( 1 - \frac{3}{2 b} \right) \left( 1 + \frac{\Delta M b^2}{8 M_c} \right)^{-1} , & \textrm{if} \, \Delta M < \frac{8 M_c}{b^2} \\
	20 \left( \frac{M_c}{1.4 b^2 + 8 M_c} \right) \left( 1 - \frac{3}{2 b} \right), & \textrm{if} \, \Delta M \geq \frac{8 M_c}{b^2} .
	\end{cases}
\end{equation}

The parameter $b$ is the ratio between the hemispheric and the polar magnetic flux. It controls the size of the mountain, with the ellipticity decreasing with increasing $b$ (see Figure 2a of \citet{2005ApJ...623.1044M}). In what follows, we refer to $b$ as the "magnetic flux ratio".

We consider a grid of values for the magnetic flux ratio.
We set a uniform grid of $\log_{10} \, b$, between $0.5$ and $3.0$, for a total of 26 bins, roughly consistent with the range of plausible values given in \citet{2005ApJ...623.1044M}, and corresponding to saturation ellipticities approximately between $\varepsilon \in [1.4 \times 10^{-9}, 7.5 \times 10^{-5}]$.

\subsection{Crust failure}
\label{sec:crust_failure}
When the spin rotation is too high for the crust to sustain it, the latter fails and breaks. \citet{2025ApJ...978L...8M} show that this can happen at spin frequencies between $40\%$ and $50\%$ of the Keplerian break-up limit. They conjecture that the reason why we see no sub-millisecond pulsars is that crust failure leaves a small deformation where the crust breaks, in turn leading to the emission of continuous gravitational waves which provide the torque necessary to halt the spin-up.

We assume crust failure to happen at $45\%$ of the Keplerian break-up spin frequency.
We calculate the latter using the following equation \citep{2013PhRvD..88d4052D}
\begin{equation}
\label{eq:keplerian_breakup}
\frac{\Omega_K}{2 \pi} = 1716 \sqrt{\frac{M/1.4M_{\odot}}{\left( R/10\textrm{km} \right)^3}} - 189 \ [\textrm{Hz}] .
\end{equation}

The deformation left by crust breakage is not predictable. 
We consider a grid of 19 ellipticities log-uniformly distributed in the range $\log \varepsilon \in [-9, -4.5]$.
The lower bound of $10^{-9}$ is chosen according to a possible indication of a minimum ellipticity among recycled pulsars \citep{2018ApJ...863L..40W}; while the upper bound of $\varepsilon \approx 3.16 \times 10^{-5}$ is chosen so that the newly formed mountain is still small enough to be long-lived \citep{2009PhRvL.102s1102H, 2022MNRAS.517.5610M}.

\section{Method and results}
\label{sec:results}

We perform a first stage of simulations without modelling the spin evolution of the neutron star for the $10332$ configurations defined in Section \ref{sec:mesa_grid} (1476 $P-M^d$ combinations for 7 accretion efficiency $\eta$ values). We select those configurations leading to systems that at some point of their evolution resemble Sco X-1 (definition in Section \ref{sec:known_params}), and re-run the simulations for these now also evolving the spin of the neutron star. We adapt the integration time-step to guarantee less than $5\%$ relative change in spin angular velocity between consecutive steps and stop our simulations when the mass ratio, which is a non-increasing monotonic function, becomes smaller than 0.28.

Since accounting for the spin evolution of the neutron star makes a MESA run slower, and given the fact that only about 4\% of all initial configurations result in a Sco X-1 compatible system, our "two-stage" simulation approach saves $\approx 90\%$ computing time in comparison with the case where we evolve the neutron star spin for all configurations.
We can proceed this way since we do not model the propeller effect, and the influence of the accretor's spin on the binary evolution is accounted for only on average through the accretion efficiency parameter, assumed to be constant.
We perform our simulation using the Atlas computing cluster at the Max Planck Institute for Gravitational Physics in Hannover.

\subsection{Sco X-1 progenitors}

Our first task is to identify the set of Sco X-1 progenitor systems.
We do not focus on this aspect of our simulations, as it has been well discussed in \citet{2021ApJ...922..174V}, and our results are largely consistent with theirs. The major difference is that we find Sco X-1 progenitors for all values of accretion efficiency $\eta$ other than for $\eta=20\%$, which is too low to result in compatible mass accretion rates.
Conversely, \citeauthor{2021ApJ...922..174V} fix the accretion efficiency exactly at $20\%$, but allow for Sco X-1 to have accretion rates as low as $0.32 \times 10^{-8}$, significantly below our lower limit of Equation~\ref{eq:cuboid}.

The number of $P-M^d$ initial configurations giving Sco X-1 progenitors depends on the accretion efficiency (see Table \ref{tab:init_configs}). The total number of $P-M^d-\eta$ initial configurations resulting in Sco X-1 progenitors is $n_{\textrm{tot}} = 460$.
The maximum compatible donor star's initial mass is slightly above $1.9 \, M_{\odot}$.
Also, the period-mass "pattern" we find resembles that of \citet{2021ApJ...922..174V} (see Figure 5 of their work) and is a manifestation of the non-trivial interplay between the evolution of the donor star and the orbital dynamics resulting from Equation \ref{eq:jdot}.
%\begin{table}
%	\centering
%	\caption{Number of $P-M^d$ initial configurations giving Sco X-1 compatible systems at different accretion efficiencies. The second column totals 460 $P-M^d-\eta$ combinations.}
%	\label{tab:init_configs}
%	\begin{tabular}{cc}
%		\toprule
%		\multirow{2}{10em}{accretion efficiency $\eta$ (per cent)} & \multirow{2}{13em}{number of Sco X-1 compatible $P-M^d$ initial configurations} \\
%		\\
%		\bottomrule
%		20& 0 \rule{0pt}{3ex} \\
%		\cmidrule(lr){1-2}
%		30& 50 \\
%		\cmidrule(lr){1-2}
%		40& 84 \\
%		\cmidrule(lr){1-2}
%		50& 86 \\
%		\cmidrule(lr){1-2}
%		60& 85 \\
%		\cmidrule(lr){1-2}
%		70& 77 \\
%		\cmidrule(lr){1-2}
%		80& 78 \\
%		\bottomrule	
%	\end{tabular}
%\end{table}
\begin{table}
	\centering
	\caption{Number of $P-M^d$ initial configurations giving Sco X-1 compatible systems at different accretion efficiencies. The second column totals 460 $P-M^d-\eta$ combinations.}
	\label{tab:init_configs}
	\begin{tabular}{cc}
		\toprule
		\multirow{2}{10em}{accretion efficiency $\eta$ (per cent)} & \multirow{2}{13em}{number of Sco X-1 compatible $P-M^d$ initial configurations} \\
		\\
		\bottomrule
		20& 0 \rule{0pt}{3ex} \\
		30& 50 \\
		40& 84 \\
		50& 86 \\
		60& 85 \\
		70& 77 \\
		80& 78 \\
		\bottomrule	
	\end{tabular}
\end{table}
%\begin{table}
%	\centering
%	\caption{Donor masses of Sco X-1 progenitor systems leading to crustal breakage. In brackets, the subset of these for which the neutron star's crust breaks during the Sco X-1 compatible phase, therefore leading to the high-frequency/loud emission of gravitational waves as discussed in Section \ref{subsec:crust_break}.}
%	\label{tab:critical_donor_masses}
%	\begin{tabular}{cl}
%		\toprule
%		accretion efficiency $\eta$ (per cent) & progenitor donor masses $M^d$ [$M_{\odot}$]\\
%		\bottomrule
%		30& -- \rule{0pt}{3ex} \\
%		\cmidrule(lr){1-2}
%		40&$M^d \geq 1.55 \ (1.57 \leq M^d \leq 1.75)$\\
%		\cmidrule(lr){1-2}
%		50&$M^d \geq 1.25 \ (1.25 \leq M^d \leq 1.37)$\\
%		\cmidrule(lr){1-2}
%		60&$M^d \geq 1.07 \ (1.07 \leq M^d \leq 1.17)$\\
%		\cmidrule(lr){1-2}
%		70&$M^d \geq 1.05 \ (1.05 \leq M^d \leq 1.1)$\\
%		\cmidrule(lr){1-2}
%		80&$M^d \geq 1.05 \, $ ($M^d \! = \! 1.05$)\\
%		\bottomrule	
%	\end{tabular}
%\end{table}
\begin{table}
	\centering
	\caption{Donor masses of Sco X-1 progenitor systems leading to crustal breakage. In brackets, the subset of these for which the neutron star's crust breaks during the Sco X-1 compatible phase, therefore leading to the high-frequency/loud emission of gravitational waves as discussed in Section \ref{subsec:crust_break}.}
	\label{tab:critical_donor_masses}
	\begin{tabular}{cl}
		\toprule
		accretion efficiency $\eta$ (per cent) & progenitor donor masses $M^d$ [$M_{\odot}$]\\
		\bottomrule
		30& -- \rule{0pt}{3ex} \\
		40&$M^d \geq 1.55 \ (1.57 \leq M^d \leq 1.75)$\\
		50&$M^d \geq 1.25 \ (1.25 \leq M^d \leq 1.37)$\\
		60&$M^d \geq 1.07 \ (1.07 \leq M^d \leq 1.17)$\\
		70&$M^d \geq 1.05 \ (1.05 \leq M^d \leq 1.1)$\\
		80&$M^d \geq 1.05 \, $ ($M^d \! = \! 1.05$)\\
		\bottomrule	
	\end{tabular}
\end{table}

\subsection{Magnetic mountains}
\label{sec:mag_detectable_configs}

To facilitate easy understanding, we examine results from one deformation mechanism at a time.
Let us start with the case of magnetic confinement of accreted matter.

The saturation ellipticity is reached almost immediately after the onset of mass transfer, and our results do not differ from the simplified treatment where the neutron star has a constant deformation since birth.
If the ellipticity is larger than $\approx 10^{-7}$ Sco X-1-like systems are typically in torque balance or very close to it ($|\dot{f}| \lesssim 10^{-12} \, \textrm{Hz} \textrm{s}^{-1}$, see Section \ref{sec:ConstraintsMagneticMountains}), the reason being that the neutron star has generally enough time to be spun-up to torque balance before the system reaches Sco X-1 compatibility. We now explain this further.

\begin{figure}
	\includegraphics[width=1.0\columnwidth]{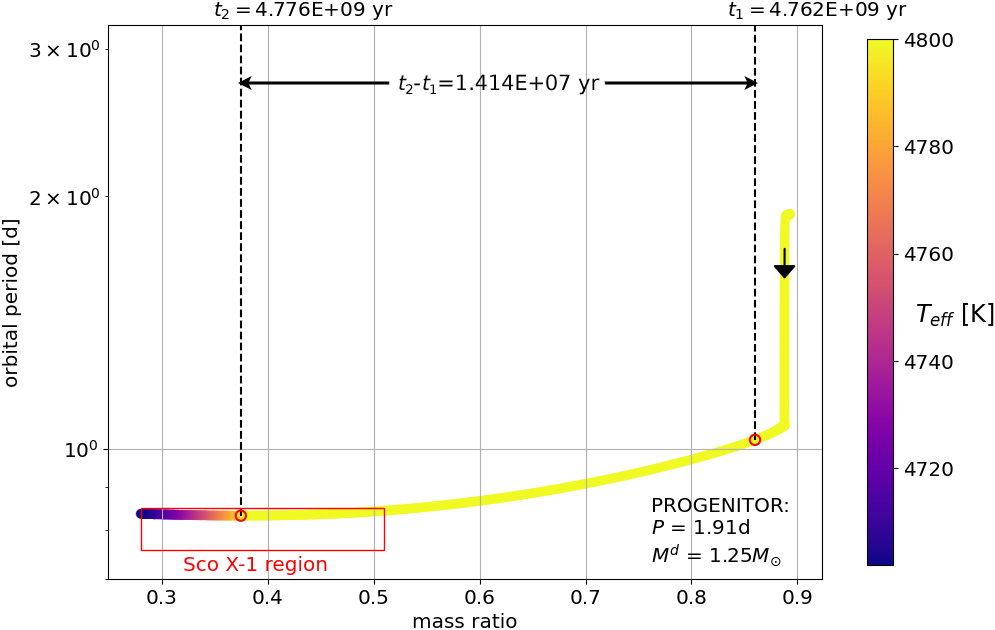}
	\caption{Orbital period and mass ratio of a Sco X-1 progenitor system as it evolves -- the arrow indicates the flow of time. The colour-code indicates the effective temperature of the donor, and temperatures non-compatible with Sco X-1 are solid yellow. The red rectangle defines the Sco X-1 compatible region. $t_1$ marks the time when the system first reaches a {\it{stable}} accretion rate of $10^{-8} \, M_{\odot} \textrm{yr}^{-1}$. $t_2$ is the earliest time when the system is Sco X-1 compatible. For this illustrative system $\eta=50$ per cent accretion efficiency. }
	\label{fig:compatibility_delay}
\end{figure}

Assuming in Equation \ref{eq:ang_mom_cons} that the torque is dominated by accretion $\tau\approx \tau_{acc}$, 
\begin{equation}
\label{eq:simplified_spinup}
\dot{\Omega} \approx \frac{\tau_{acc}}{I} \approx \frac{\dot{M}_{a} (G M R)^{1/2}}{I}.
\end{equation}
Taking the right-hand side as constant and integrating between $\Omega\approx 0$ at $t=0$ and $\Omega_{tb}$ at $\Delta t$, yields 
\begin{equation}
\label{eq:deltat_tb}
\Delta t = \frac{2 \pi \, \nu_{tb} \, I}{\dot{M}_{a} (G M R)^{1/2}} ,
\end{equation}
where $\nu_{tb}$ is the spin frequency at torque balance. From the approximate form of the torque balance condition 
\begin{equation}
\label{eq:approx_tb}
\tau_{acc} = - \tau_{cgw}
\end{equation}
we find
\begin{equation}
\label{eq:omega_tb}
\begin{split}
\nu_{tb} = 346.2 \, & \left( \frac{M}{1.4 M_{\odot}} \frac{R}{10\textrm{km}} \right)^{1/10} \left( \frac{\dot{M}_a}{10^{-8} M_{\odot} \cdot \textrm{yr}^{-1}} \right)^{1/5} \\
& \left( \frac{I}{10^{45} \textrm{g} \cdot \textrm{cm}^2} \frac{\varepsilon}{10^{-7}} \right)^{-2/5} \textrm{Hz}
\end{split}
\end{equation}
If we take the quantities in brackets of the Equation above to be equal to 1, from Equation \ref{eq:deltat_tb} we find
\begin{equation}
\label{eq:deltat_tb_estimate}
\Delta t = 1.3 \times 10^4 \, \varepsilon^{-2/5} \, \textrm{yr} .
\end{equation}
So, a neutron star with $\varepsilon \gtrsim 10^{-7}$ accreting at Sco X-1 typical accretion rates, takes $\lesssim 8$ million years to reach torque balance. 
Figure \ref{fig:compatibility_delay} shows a system that has more than 14 million years of stable accretion at $\dot{M}_a \gtrsim 10^{-8} \, M_{\odot} \textrm{yr}^{-1}$ before it is Sco X-1-like. This means that if such system has an ellipticity of at least $\varepsilon \approx 10^{-7}$, when it reaches the Sco X-1-like configurations, it is in torque balance.
With different progenitors and different accretion efficiencies, the stable accretion phase before the system is Sco X-1-like, changes in duration; however, it is never smaller than a few million years.

\subsubsection{Effects of accretion efficiency and mass of the donor star}
\label{sec:effects_acceff_donormass}

\begin{figure*}
	\includegraphics[width=0.9\linewidth]{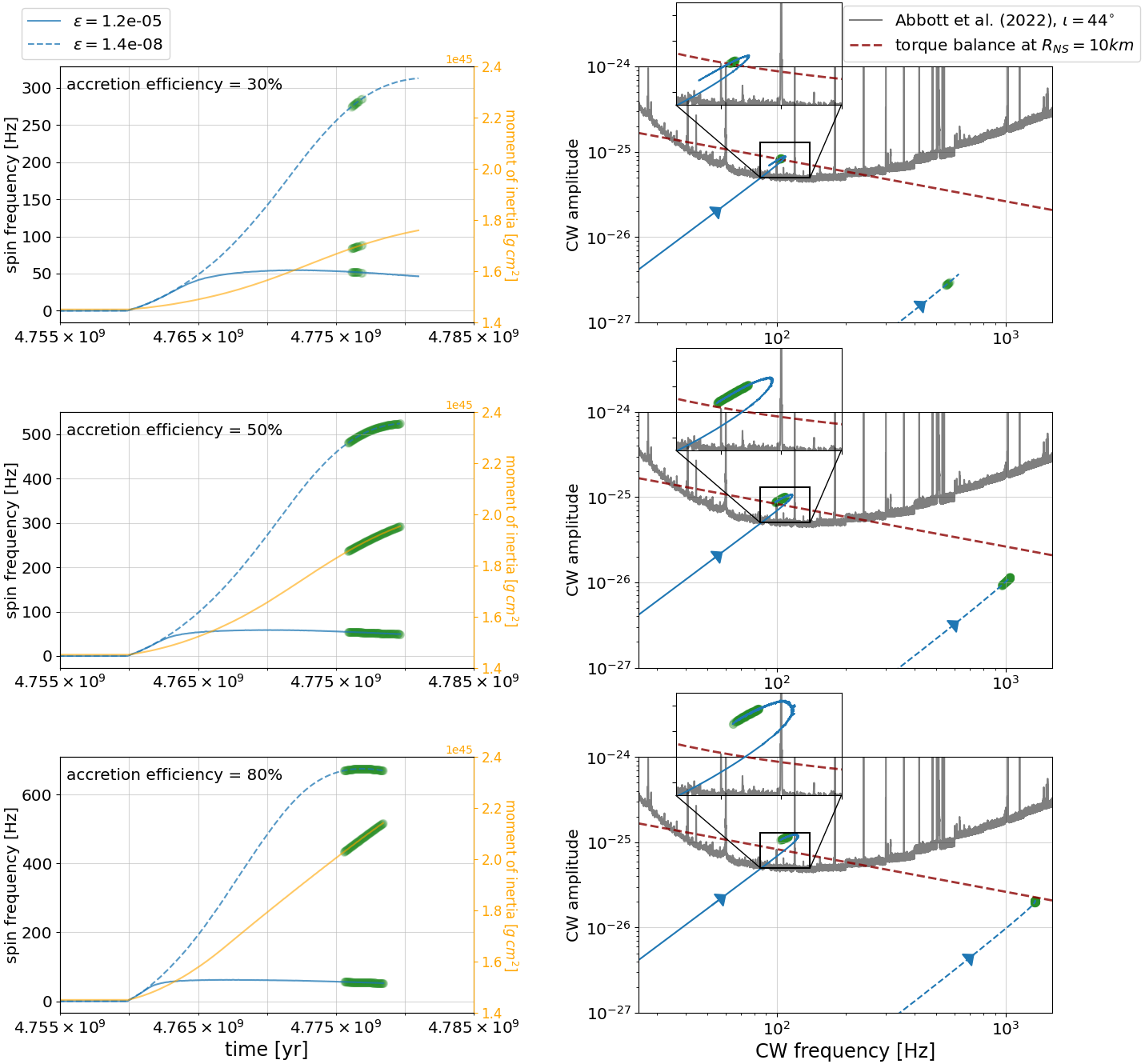}
	\caption{The three rows of plots differ for the value of the accretion efficiency: 30\%, 50\% and 80\%, and refer to the same progenitor system defined by: $P=1.91 \, \textrm{d}, M^{d}=1.25 \, M_{\odot}$. On the left we have the spin frequency evolution of the neutron star in Sco X-1 in the case of high deformation (solid blue curve) and small deformation (dashed blue curve), in both cases due to magnetic confinement of accreted matter. The twin $y$ axis shows the moment of inertia as a function of time (orange curve). We highlight the evolutionary stage where the system resembles Sco X-1 with scatter markers superimposed on the curves (in this example, in green). On the right, we plot the gravitational wave amplitude $h_0(f)$ emitted by the highly and weakly deformed neutron star (same as the left-side plots). 
We compare it with the most recent gravitational wave amplitude upper limits $h_0^{ul}$ (\citet{2022ApJ...941L..30A}, solid grey curve), as well as with the nominal torque-balance condition introduced in Section \ref{sec:torques}.
The $h_0(f)$ curves present a "turning point" in the case of high ellipticity: very early on, the system reaches torque balance, so the frequency stops increasing. However, since the mass accretion rate $\dot{M}_\textrm{a}$ decreases over time, the accretion torque also decreases, and the neutron star begins to slowly spin down. Even though the moment of inertia $I$ increases, since $h_0 \propto I \, f^2$, overall $h_0$ decreases, creating the ``turning point".
}
	\label{fig:evolution_example}
\end{figure*}

A highly deformed neutron star reaches torque balance quickly and at quite low spin frequencies. Spin-up is halted very early in the accretion, hence the accretion time scale is largely irrelevant (solid lines in the left-hand-side plots of Figure \ref{fig:evolution_example}). Conversely, a weakly deformed neutron star needs higher frequencies to balance the accretion torque, potentially facing a much longer spin-up phase. The latter is halted either because torque balance is effectively reached, or because mass accretion has stopped or significantly decreased. However, the efficiency of the spin-up and the duration of the accretion phase vary with different progenitor-accretion efficiency combinations, making predictions of the spin evolution hard to make. We address this difficulty with a numerical approach.

With increased accretion efficiency, we essentially have two effects on the neutron star: a stronger accretion torque and a heavier neutron star\footnote{Heavier compared to a system with the same $q$, same progenitor but accreting mass at a lower efficiency.}.
A stronger accretion torque means a more efficient spin-up. This is why the weakly deformed Sco X-1 neutron star of Figure \ref{fig:evolution_example} spins faster as accretion efficiency increases.
As for a heavier neutron star, it generally yields a larger accretion torque and a larger gravitational wave torque (Equations \ref{eq:acc_torque} and \ref{eq:cgw_torque}), so in the end the torque-balance frequency does not change very much with the accretion efficiency.
What changes appreciably is instead the gravitational wave amplitude $h_0 \propto I$, hence gravitational waves are stronger at higher accretion efficiencies.
We see this effect in the insets of the right-hand-side plots of Figure \ref{fig:evolution_example}.

\begin{figure*}
	\includegraphics[width=0.9\linewidth]{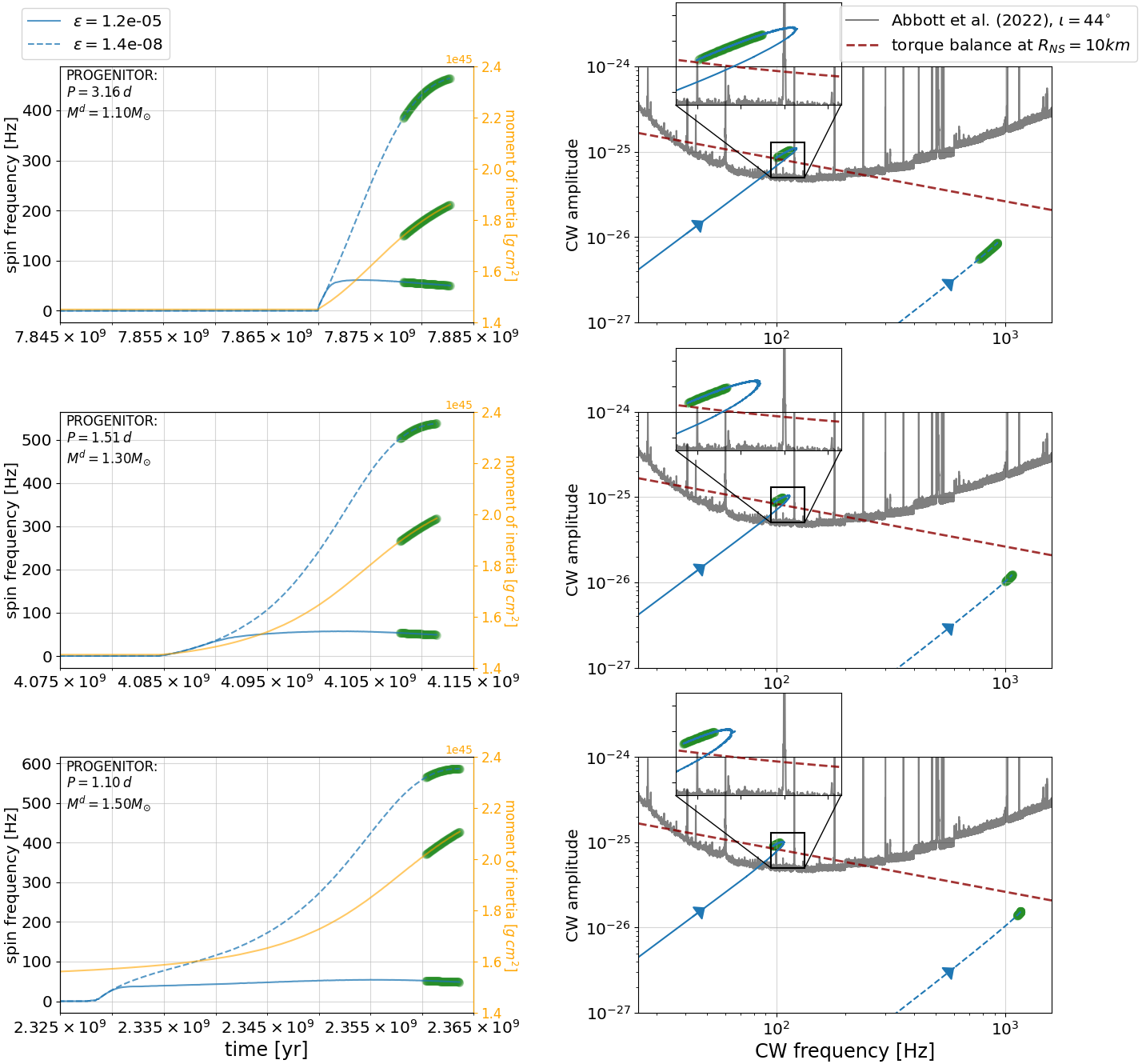}
	\caption{Same as Figure \ref{fig:evolution_example} but in each row of plots we vary the progenitor system, maintaining the accretion efficiency constant at $\eta=50\%$.
	Generally, the spin-up starts right at the onset of mass transfer, so we can compare the spin-evolution phases to see how the length of the mass transfer phase increases as the mass of the donor star increases.
	Notice the left plot in the third row: at $t=2.325 \times 10^9 \, \textrm{yr}$ the spin-up has not yet started but the moment of inertia is already above the initial value of $\approx 1.45 \times 10^{45} \textrm{g} \textrm{cm}^2$. This is because, for a certain period of time, the mass transfer rate is non-zero but the disc-sustained condition of Equation \ref{eq:dubuis} does not hold and the accretion torque is zero; the neutron star can however accrete mass, increasing its moment of inertia.
	Progenitor donor stars with masses greater than $1.5 \, M_{\odot}$ are omitted from the figure because their multiple accretion episodes introduce complexities that hinder the intended interpretability of the visualization. See \citet{2025AA...693A.314M} for an explanation of why multiple Roche lobe overflow phases happen in systems with heavy donors.}
	\label{fig:evols_donors}
\end{figure*}

Figure \ref{fig:evols_donors} shows that for $\eta = 50\%$, $M^d \leq 1.5 \, M_{\odot}$ and $\varepsilon \approx 10^{-8}$, a Sco X-1-like neutron star is not in torque-balance: the gravitational wave torque is not large enough to balance the accretion torque for the values of the mass ratio that are compatible with Sco X-1. In other words, Sco X-1-like configurations spin too slowly for torque balance. 
This changes as $M^d$ increases, because Sco X-1-compatible mass ratios are now reached at a higher spin frequency (the neutron star is being spun-up for longer, since it needs to accrete more mass to reach the same Sco X-1-like $q$). At these higher frequencies the gravitational wave torque is much larger and there exists a value of $M^d$ for which, even for very small ellipticities, Sco X-1-like configurations can be in torque balance. 
As $M^d$ increases the gravitational wave torque increases not only through the higher spin frequency, but also, and quadratically, through the mass of the neutron star. So at fixed mass ratio (Sco X-1-like), heavier donors produce heavier neutron stars for which the accretion torque is balanced at comparatively {\it{lower}} frequencies than for lighter donors.
This latter effect is clearly visible in the insets of Figure \ref{fig:evols_donors} where torque balance is reached at progressively smaller spin frequencies as the donor star is heavier.

\subsubsection{Constraints on magnetic mountains}
\label{sec:ConstraintsMagneticMountains}

For magnetic mountains, the magnitude of the ellipticity depends on the magnetic flux ratio $b$, which can hence be constrained based on null-detection results. We first review the most common way to assess detectability in the absence of a population study, and then present our results.

\underline{Standard detectability:} the most frequently used way to assess detectability of a signal at a certain frequency by a given search, is to assume that Sco X-1 is in torque balance (Equation~\ref{eq:torque_balance_general}), so that the gravitational wave frequencies $f$ at which the signal is detectable are
\begin{equation}
\label{eq:detectability}
f ~ \mid ~ h_0^{tb}(f) \geq h_0^{ul}(f),
\end{equation}
where $h_0^{tb}$ is the torque balance amplitude and $h_0^{ul}$ is the observational upper limit value. The $h_0^{ul}$ can then be translated in a lower limit on $b$, using Equation~\ref{eq:melatospayne_adapted} with $\Delta M \geq 8 M_c / b^2$. 

One often finds the torque balance amplitude computed from Equation~\ref{eq:torque_balance_general} by assuming nominal values for $M, R_i$ and $\dot{M}_a$, most often taken equal to the reference scaling ones used in our equation. With this convention and considering the most recent gravitational wave amplitude upper limit results from \citet{2022ApJ...941L..30A}, we find that the detectability region ($h_0^{tb} \geq h_0^{ul}$) falls roughly between 40 and 200 Hz. In this region, the null detection results yield $b \gtrsim 18$, corresponding to $\varepsilon \leq 3.9 \times 10^{-6}$.

\underline{Direct determination:}  based on our results we can determine the detectability directly:  a signal stemming from a simulated system is detectable if its frequency and frequency time derivative fall within the parameter space covered by \citet{2022ApJ...941L..30A} and if its amplitude is greater than amplitude upper limits at the signal's frequency.

The frequency variability of Sco X-1 can be both short-term -- with timescales ranging from months to thousands of years -- and long-term -- with timescales ranging from hundreds of thousands to millions of years.
Our study is not suited to forecast short-term variability on Sco X-1.
Firstly, because we do not model very short-term effects such as spin-wandering \citep{2018PhRvD..97d3016M} characterised by timescales of the order of days to months.
Secondly, the mass transfer rates of systems simulated with MESA present numerical noise affecting the spin frequency time derivative on timescales $\lesssim 10^{5} \, \textrm{yr}$, therefore impacting the higher end of timescales pertaining to short-term variability.

What we can do here is to outline the long-term variability of simulated signals, which is still relevant from a search perspective.
We calculate the long-term frequency time derivative of a signal as follows
\begin{equation}
\label{eq:fdot_lt}
\dot{f}^{\textrm{lt}} = \frac{f_{\textrm{out}} - f_{\textrm{in}}}{t_{\textrm{out}} - t_{\textrm{in}}}
\end{equation}
where $f_{\textrm{in}}$ is the frequency of the gravitational wave signal emitted when the system first resembles Sco X-1 at $t_{\textrm{in}}$, and $f_{\textrm{out}}$ the frequency when it last resembles Sco X-1 at $t_{\textrm{out}}$, approximating this to be constant throughout the entire Sco X-1 compatible phase (See discussion in Section \ref{sec:cav_and_crit}).

The $\dot{f}$ range covered by \citet{2022ApJ...941L..30A}, is not given in the paper.
We estimate the latter by using Equation 86 of \citet{2015PhRvD..91j2003L}, with $t_s=T_\textrm{obs} $, consistent with having assumed that the timescale during which the signal varies is equal to the total observing time $T_{\textrm{obs}}$:
\begin{equation}
\label{eq:max_fdot}
\mid \dot{f}_{\textrm{max}} \mid = \frac{\sqrt{2}}{ T_\textrm{obs} } \Delta f,
\end{equation}
where $\Delta f$ is the frequency resolution. 
From Table 3 of \citet{2022ApJ...941L..30A} the frequency resolution $\Delta f \geq 5.3 \times 10^{-5} \, \textrm{Hz}$, therefore using $T_{\textrm{obs}} = 3 \times 10^7 \, \textrm{s}$ which is the duration of the LIGO O3 run, we find $\mid \dot{f}_{\textrm{max}} \mid \geq 2.4 \times 10^{-12} \, \textrm{Hz} \textrm{s}^{-1}$.

All the simulated signals emitting gravitational waves at the current detectable level have $-6 \times 10^{-13} \leq \dot{f}^{\textrm{lt}} / [\textrm{Hz} \textrm{s}^{-1}] \leq 2.8 \times 10^{-13}$ and frequencies much below 1600 Hz, therefore they all lie within the parameter space coverage of \citet{2022ApJ...941L..30A}.

The highest frequency at which we find detectable signals grows with the accretion efficiency: for $\eta=30\%$ the highest frequency is $\approx190 \, \textrm{Hz}$, at $\eta = 80\%$ it is $361 \, \textrm{Hz}$.
As already explained in Section \ref{sec:effects_acceff_donormass}, a larger accretion efficiency produces a stronger accretion torque and a heavier neutron star which result in stronger gravitational waves -- see insets of Figure \ref{fig:evolution_example} .

Detectability established by direct comparison of the upper limits with our simulated Sco X-1 population yields detectable signals at frequencies higher than $\approx$ 200 Hz, which is instead the highest currently detectable frequency based on nominal torque balance. The primary reason is that in nominal torque balance $R_i=10$ km, whereas in our simulations we find $17 \leq R_i / [\textrm{km}] \leq 30$. As explained at the beginning of this Section, the nominal torque balance condition is often used as a standard benchmark to gauge to what extent null search results are astrophysically relevant. Our numerical results confirm what is already understood in searches, namely that the gravitational wave amplitude calculated assuming the nominal torque balance condition is a conservative estimation of the strength of the gravitational waves emitted by Sco X-1, assuming this is in torque balance.

We can summarise as follows: at $\eta = 30\%$ our results yield $\varepsilon < 2.1 \times 10^{-6}$, corresponding to $b > 25$ for frequencies below $\approx 200 \, \textrm{Hz}$. 
More stringent constraints -- $\varepsilon < 8.7 \times 10^{-7}$ ($b > 40$) -- can be set for Sco X-1 progenitors with highly efficient mass transfer $\eta \geq 70\%$, for frequencies below $\approx 360 \, \textrm{Hz}$.

\subsubsection{Third generation detectors}
\label{sec:3g_det_magnetic}
As the sensitivity of the detectors increases, signals stemming from neutron stars with increasingly smaller deformations come into reach.
This means that the effects of accretion efficiency and mass of the progenitor donor star treated in the previous subsection and pertaining to weakly deformed neutron stars, begin to play a role in the observations.

Let us consider third-generation ground-based detectors. Following the procedure already adopted and amply discussed in \citet{2023ApJ...952..123P} and \citet{2025MNRAS.540.1006P}, we imagine to replicate the search done by \citet{2022ApJ...941L..30A}, but on more sensitive data.
This search reaches a sensitivity depth \citep{2015PhRvD..91f4007B} of $\mathcal{D} \approx 62 \, \textrm{Hz}^{-1/2}$ using both LIGO detectors.
The amplitude of the weakest detectable signal at frequency $f$ is therefore\footnote{See Section 2.1 of \citet{2025MNRAS.540.1006P} for a detailed discussion and derivation.}
\begin{equation}
\label{eq:h0_future}
h_0^{3G}(f) = \frac{\sqrt{S_h^{3G}(f)}}{62 \, [\textrm{Hz}]^{-1/2}} \sqrt{\frac{2}{N}},
\end{equation}
where "$3G$" refers to quantities pertaining to third generation detectors, $\sqrt{S_h}$ is the amplitude spectral density and the factor $\sqrt{2/N}$ accounts for $N$ possible independent data streams available.
We find virtually no difference between results obtained considering Cosmic Explorer in the baseline 40 km single detector design ($N=1$, \citet{evans2023cosmicexplorersubmissionnsf, ce_asd_citation}), and Einstein Telescope in the triangular configuration ($N=3$, \citet{2011CQGra..28i4013H, 2020JCAP...03..050M}). For ease of presentation and treatment, we show results obtained considering Cosmic Explorer only, as representative of third generation detectors.

For $3G$ detectors the detectability criterion is the same as the one adopted for current sensitivity, that is a simulated signal at frequency $f$ with amplitude $h_0(f)$ must have: $h_0(f) \geq h_0^{3G}(f)$ and $\mid \dot{f}^{\textrm{lt}} \mid \leq 2.4 \times 10^{-12} \, \textrm{Hz} \textrm{s}^{-1}$.

All the simulated signals emitting gravitational waves at $3G$ detectable level have $-6.1 \times 10^{-13} \leq \dot{f}^{\textrm{lt}} \leq 2.2 \times 10^{-12}$.
We can conclude that, as far as magnetic mountains are concerned, the long-term spin frequency evolution does not impact the sensitivity of gravitational wave searches like \citet{2022ApJ...941L..30A}, at least for observation times of $\approx 1 \, \textrm{yr}$.

\begin{figure}
	\includegraphics[width=1.0\columnwidth]{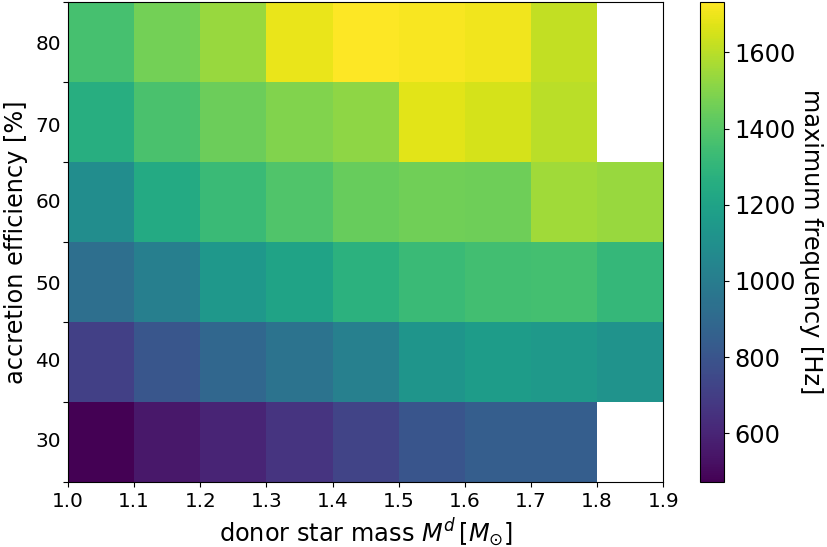}
	\caption{The $M^d-\eta$ space colour-coded with the maximum frequency of continuous gravitational waves detectable by $3G$ detectors. If we do not find Sco X-1 compatible configurations the bin is left empty (white). The pattern shown in this plot can also be appreciated by looking at the summary Figure \ref{fig:h0_vs_f_mag} in appendix \ref{sec:appendix_figs}.}
	\label{fig:eta_donormass_frequency_map}
\end{figure}

The most prominent effect that progenitor-accretion efficiency combinations have on the 3G-detectability of continuous waves from Sco X-1 relates to the maximum frequency at which detectable signals are found.
Figure \ref{fig:eta_donormass_frequency_map} shows the maximum frequency of detectable signals as progenitor and accretion efficiency vary, featuring effects already discussed in Section \ref{sec:effects_acceff_donormass}: at fixed donor star mass, as the accretion efficiency increases, the frequency of detectable signals increases.
The same happens at fixed accretion efficiency as the mass of the progenitor donor star increases up to $\approx 1.5 - 1.8 \, M_{\odot}$, depending on the value of $\eta$, to then slightly decrease as $M^d$ increases further. 
This latter fact is a manifestation of how even the Sco X-1 configurations with the smallest detectable deformations reach torque balance in systems with heavy donors and a highly efficient mass transfer.

\subsubsection{Sco X-1 detection probability}
\label{sec:sco_det_prob_mag}

It is possible to measure how likely a detection of continuous gravitational waves from Sco X-1 is in the magnetic mountain scenario.
In Section \ref{sec:mesa_grid} and \ref{sec:mag_conf} we have described how we divide the parameter space linearly along the 4 dimensions $(\log P-M^{\textrm{d}}-\eta-\log b)$ in $36\times 41\times 7\times 26 = 268\,632$ bins. We consider only the $n_{\textrm{tot}}$ $(P-M^{\textrm{d}}-\eta)$ combinations that can reproduce Sco X-1, and get $N = n_{\textrm{tot}} \times 26 = 11960$ bins in $(\log P-M^{\textrm{d}}-\eta-\log b)$. 
We can assign the probability $p^j$ of the $j^{th}$ bin to be Sco X-1 progenitor\footnote{We are here extending the concept of "progenitor" also to the dimensions defined by $\eta$ and $\log \, b$.} to be
\begin{equation}
\label{eq:p_j}
p^j = \frac{t_{res}^j}{\sum_{i=1}^{i=N} t_{res}^i}
\end{equation}
where $t_{res}^i$ is the time the $i^{th}$ bin spends resembling Sco X-1.
Now, the probability $p_{det}^j$ of the $j^{th}$ bin to be Sco X-1 progenitor {\it{and}} to emit a detectable gravitational wave is
\begin{equation}
\label{eq:p_det^j}
p_{det}^j = p^j \cdot \frac{t_{det}^j}{t_{res}^j} = \frac{t_{det}^j}{\sum_{i=1}^{i=N} t_{res}^i}
\end{equation}
where $t_{det}^j$ is the total time during which the $j^{th}$ bin emits a detectable gravitational wave.
Therefore
\begin{equation}
\label{eq:p} 
p:=\sum_{i=1}^{i=N} p_{det}^i
\end{equation}
provides a measure of the likelihood of Sco X-1 being detectable.

For $3G$ detectors we find $p^{3G} \approx 0.82$.
For {\it{current}} detector sensitivity, $p$ decreases by a factor of $\approx 2$. In Appendix \ref{sec:appendix_figs} we present the summary Figure \ref{fig:h0_vs_f_mag} for this deformation case.

We provide the $t_{res}^i$ and $t_{det}^i$ for magnetic mountains and for crustal breakage (Section~\ref{sec:det_prob_crust_break}), for this and the next generation of ground-based detectors, in machine-readable format as \citep{supplementary_material}.

\subsection{Crustal breakage}
\label{subsec:crust_break}

\begin{figure*}
	\includegraphics[width=1.0\linewidth]{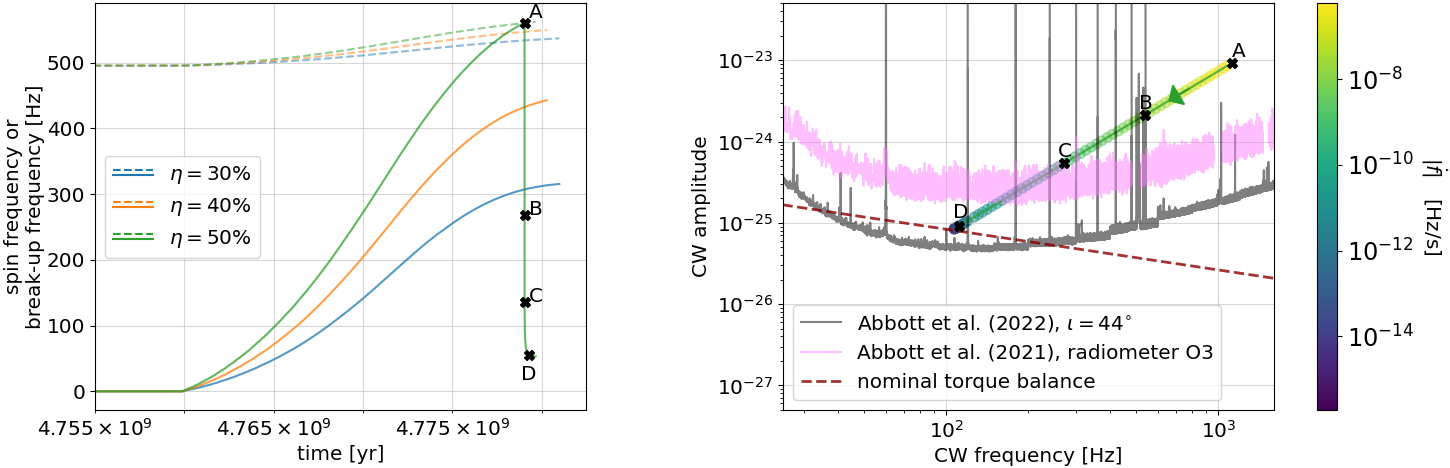}
	\caption{Spin evolution (solid lines, left plot) and concurrent gravitational wave emission (right plot) of a Sco X-1 neutron star going through crustal failure, which leaves behind a residual ellipticity of $\varepsilon = 10^{-5}$, for three values of the accretion efficiency: 30 per cent, 40 per cent and 50 per cent; among these, only $\eta=50$ per cent leads to crustal failure therefore to continuous gravitational wave emission. The dashed lines on the left plot represent $45$ per cent of the Keplerian break-up spin frequency as a function of time. The gravitational wave "trace" of the right plot is colour-coded with the instantaneous gravitational wave frequency time derivative.
	It takes $\approx 250 \, \textrm{yr}$ to spin-down from A, the point of crustal breakage, to B, where spin frequency is half of the break-up frequency. It takes further $\approx 3500 \, \textrm{yr}$ to reach C, where the spin frequency is reduced by another factor of 2. In D, reached from C in $\approx 200 \, 000 \, \textrm{yr}$, the neutron star is in torque balance, persisting there for slightly less than a million years before leaving the Sco X-1 compatibility state. 
	The progenitor system considered here is the same as the one considered in Figure \ref{fig:evolution_example}.}
	\label{fig:break_evolution}
\end{figure*}

We next present results for the case where the crust fails to sustain centrifugal stresses and breaks, leaving behind a residual deformation $\varepsilon$ (Section~\ref{sec:crust_failure}). Gravitational wave emission only begins after the breakage. This is illustrated in Figure \ref{fig:break_evolution}, where we can observe that the large assumed ellipticity $\varepsilon = 10^{-5}$ is responsible for a dramatic change in the spin frequency evolution after the breakage. In our simulation we consider a range of $\varepsilon \in [10^{-9}, 3.16 \times 10^{-5}]$.

After crustal breakage, the spin frequency reduces rapidly if $\varepsilon \gtrsim 5 \times 10^{-8}$, reaching torque balance in approximately $150\,000 \, ( \varepsilon / 10^{-5} )^{-2/5} \, \textrm{yr}$, where it persists for a few million years, depending on the system, before leaving the Sco X-1 compatibility state. 
In Figure \ref{fig:break_evolution} we break down the spin evolution of the neutron star to visualise the rapid spin-down in the case of high residual deformation. 

\begin{figure}
	\includegraphics[width=1.0\columnwidth]{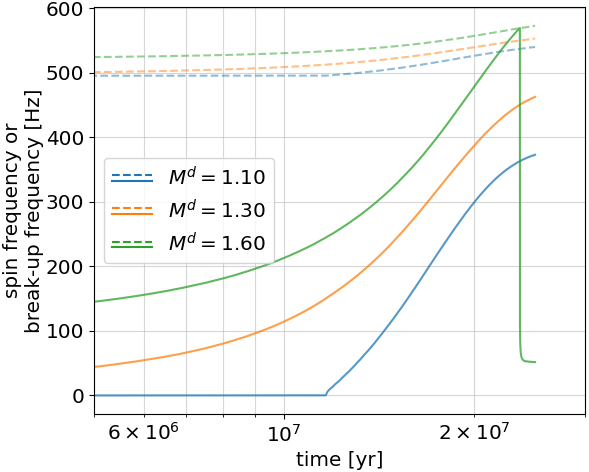}
	\caption{Same as left-hand-side of \ref{fig:break_evolution} but where we vary the progenitor system and keep constant the accretion efficiency at $\eta=40$ per cent. Also, since the onset of mass accretion of different progenitor systems happens at different times, the x-axis is not the time since birth, but represents the last 25 million years of evolution for all three systems before the end of the simulation.}
	\label{fig:break_evolution_2}
\end{figure}

The breakage, and hence the gravitational wave emission, is predicated upon the neutron star reaching the angular velocity $\Omega^{\textrm{br}} = 0.45 \times \Omega^K$. This introduces a strong dependency on the accretion efficiency because a larger accretion efficiency produces a more efficient spin-up, as discussed in Section \ref{sec:effects_acceff_donormass}.
We find that for $\eta \leq 30\%$ no binary configuration compatible with Sco X-1 manages to spin the neutron star up to $\Omega^{\textrm{br}}$. For $\eta = 40\%$ crustal failure only happens for Sco X-1 systems with donor mass exceeding $1.55 \, M_{\odot}$ (see Figures \ref{fig:break_evolution_2} and \ref{fig:h0_vs_f_break}). As the accretion efficiency increases, more and more Sco X-1 systems present crustal breakage, as lower mass donors are able to sufficiently spin up the star. For  $\eta \geq 70\%$, all progenitors lead to crustal breakage (see Table \ref{tab:critical_donor_masses}).

One of the distinct features of this deformation mechanism is the strength of emitted gravitational waves, which reaches amplitudes almost two orders of magnitude higher than the magnetic mountain case, as shown in Figure~\ref{fig:h0_vs_f_break}, which serves as a summary figure for this deformation case. This is due to the fact that the gravitational wave emission starts at $f^{\textrm{br}} = \Omega^{\textrm{br}} / \pi$ which is high (above $\approx 1000$ Hz) and the gravitational wave amplitude is quadratic in $f$ (Equation~\ref{eq:h0}). If the system is in the Sco X-1 compatible phase when the crust breaks, this means that the entire signal could be observed, as it sweeps from $f^{\textrm{br}}$ to the much lower torque balance frequency. 
It should however be noted that the high-frequency/loud gravitational wave phase is very short (tens of thousands of years) compared to the low-frequency/weak gravitational wave phase (hundreds of thousands to millions of years), making it unlikely to observe such signal during its loudest phases. The rapid spindown in the loud phase also falls outside the range covered by the most sensitive searches like \citet{2022ApJ...941L..30A}. It is nonetheless covered by the radiometer searches, whose sensitivity in \citet{2021PhRvD.104b2005A} would be sufficient to detect all signals in the high-frequency/loud gravitational wave phase if the ellipticity after crustal breakage is $\varepsilon \geq 10^{-6}$, highlighting the importance of robust detection techniques.

We point out the relevance of this deformation mechanism also when we extend our horizon to extra-Galactic objects. Indeed, the strength of a continuous gravitational wave linearly decreases with distance (see Equation \ref{eq:h0}), but the number of galaxies, and therefore the chances to find a Sco X-1-like system just going through crustal failure, increases with the third power of the distance. 
Making an estimation of our chances to detect a Sco X-1-like extra-Galactic system while it goes through crustal breakage is however a work on its own, and we reserve the right to investigate this matter further in the near future.

\subsubsection{Sco X-1 detection probability in the case of crustal breakage}
\label{sec:det_prob_crust_break}

We estimate the probabilities $p_r$ and $p_t$ of detecting continuous gravitational waves from Sco X-1 crustal failure by respectively a radiometer search like \citep{2021PhRvD.104b2005A} and by a template-search like \citep{2022ApJ...941L..30A}.

Let us start estimating $p_{r}$.
We adopt the same detectability criterion as in Section \ref{sec:ConstraintsMagneticMountains}. The radiometer search is in principle sensitive to any frequency time derivative and it covers almost the entire frequency range available, $20 \leq f / [\textrm{Hz}] \leq 1726$, with a few data gaps around glitches and known instrumental lines at $\approx 500, 1000 \, \textrm{and} \, 1500 \, \textrm{Hz}$. Based on current results we estimate a radiometer sensitivity depth of $\mathcal{D}_r = 14~\textrm{Hz}^{-1/2}$ and assume this holds throughout the $3G$ detector band. 

The parameter space binning is here uniform in $\log P-M^d-\eta-\log \varepsilon$.
A huge portion of this space, measuring $30.6\%$ of the total, simply does not lead to gravitational wave emission.
For current radiometer searches, we estimate $p_{r}^{LIGO} = 2.2 \times 10^{-4}$, meaning that current chances to detect a signal from Sco X-1 going through crustal failure are less than $0.03\%$.
For searches based on $3G$ detectors data, we estimate $p_{r}^{3G} = 0.41$. Such a huge difference between current and $3G$ results can be understood by noting that the sensitivity of current radiometer searches allows for the detection of signals only in the high-frequency/loud phase, which are fast-evolving and also happen only for a very small subset of Sco X-1 configurations.
Future radiometer searches will, in principle, allow for the detection of the entirety of gravitational wave signals in both their high-frequency/loud phase and low-frequency/weak phase (see Figure \ref{fig:h0_vs_f_break}).

We conclude estimating $p_{t}$.
The detection criterion used in Sections \ref{sec:ConstraintsMagneticMountains} and \ref{sec:3g_det_magnetic} applies here.
We discard the portion of simulated signals where $\dot{f} < -2.4 \times 10^{-12} \, \textrm{Hz} \textrm{s}^{-1}$, i.e. where the spin-down is too high for the signal to be detectable, and calculate $t_{det}^i$ of Equation \ref{eq:p_det^j} accordingly.
We find: $p_{t}^{LIGO} = 0.21$ and $p_{t}^{3G} = 0.45$.

\section{Discussion}
\label{sec:discussion}
We have simulated the evolution of Sco X-1 in all its parts, focusing on the emission of continuous gravitational waves by the accreting neutron star. 
We have adopted the reverse population synthesis technique introduced by \citet{2021ApJ...922..174V}, extending the list of free parameters to include the accretion efficiency and a parameter related to the neutron star deformation.
We assume the neutron star has a non-axisymmetric mass distribution, due to the following mechanisms:
\begin{enumerate}
\item{Magnetic confinement of accreted matter}
\item{Leftover deformation after crustal breakage.}
\end{enumerate}
We study the detectability and features of the resulting gravitational signals, in relation to the different possible parameter combinations.

\begin{description}

\item[\textbf{Magnetic confinement of accreted matter:}] 
Neutron stars with deformations $\varepsilon \gtrsim 10^{-7}$ can reach torque balance before the system starts to resemble Sco X-1, regardless of progenitor or mass accretion efficiency, and are in a Sco X-1-like state while in torque balance for half a million - few million years. The effects of progenitor or mass accretion efficiency are mild, and yield slightly stronger signals as accretion efficiency increases, and slightly lower torque-balance frequencies as the mass of the donor star increases.

The long-term spin frequency time derivative is within the search range of current continuous gravitational wave searches, at least for observing periods of $\approx 1 \, \textrm{yr}$.

The highest frequency at which we find currently detectable gravitational waves depends on the accretion efficiency, going from $f \approx 200 \, \textrm{Hz}$ when $\eta = 30\%$ up to $f \approx 360 \, \textrm{Hz}$ for $\eta \geq 70\%$; the latter value is significantly higher than what can be forecasted by means of the nominal torque-balance condition.
At current sensitivity, and assuming the average accretion efficiency of Sco X-1 to be $\eta \approx 30 \%$, we can constrain the neutron star in Sco X-1 to have $\varepsilon < 2.1 \times 10^{-6}$, which corresponds to magnetic flux ratios $b > 25$.
Slightly more favourable is the scenario of Sco X-1 accreting at $\eta \geq 70 \%$, in which case we obtain the constraints: $\varepsilon < 8.7 \times 10^{-7}$, corresponding to $b > 40$.

If the built-up mountain is $\varepsilon < 10^{-7}$ we do not expect detections with the current generation of gravitational wave detectors. Signals detectable by third generation detectors have frequencies that increase with accretion efficiency and mass of progenitor donor star, with the highest detectable frequency in the broad range $[600, 1700] \, \textrm{Hz}$.

Assuming an approximately log-uniform distribution of ellipticities $\in [10^{-9}, 7 \times 10^{-5}]$ the detection probability is 43\% with current detectors, while it increases to $82\%$ for $3G$ detectors. 
We provide data so the interested reader can estimate the detection probability for narrower ranges of parameters.

\item[\textbf{Leftover deformation from crustal breakage:}] 
accretion efficiency level and donor mass hugely impact gravitational wave emission: 
at $30\%$ accretion efficiency, we have no crustal breakage, therefore no gravitational wave emission.
At higher accretion efficiencies, only a subset of Sco X-1 progenitors lead to the failure of the neutron star's crust.
In particular, for efficiencies between $40\%$ and $60\%$, we find a lower bound on the mass of the progenitor donor star leading the neutron star's crust to break. This lower bound goes from $1.55 \, M_{\odot}$ to $1.07 \, M_{\odot}$, decreasing with increasing efficiency of the mass transfer.

For a few progenitor-accretion efficiency combinations the crust of the neutron star breaks while the system is already Sco X-1-like, so we can, in principle, observe the entire signal sweeping from $\approx 1000 \, \textrm{Hz}$ to the torque-balance gravitational wave frequency. This high-frequency/loud gravitational wave phase cannot be detected by template-based searches, since the spin-down during this phase is too high.
Radiometer searches or hidden Markov model searches \citep{2018PhRvD..97d3013S, 2021PhRvD.104d2003M} are better suited to detect signals emitted during this phase. However, current detection probability is hindered by the fact that the detectable phase is very short-lasting and by the fact that the number of Sco X-1 configurations for which this happens is only a few. We find virtually no chances to detect such signals now. 

Detection prospects with third-generation detectors are much brighter: not only is a much larger portion of the sweeping signal detectable, but also emission at torque balance, at least for the large majority of the possible signals. We estimate a probability of $41\%$ to detect gravitational waves from Sco X-1 by means of radiometer searches.
If the leftover post-breakage ellipticity remains unchanged, the neutron star will keep emitting continuous gravitational waves in torque balance, which could be detected by the more sensitive template-based searches. We estimate the detection probability for these to be $45\%$.

We note that these estimates rely on the fact that the crust breaks when $\Omega^{\textrm{br}}/2\pi \approx 550 \, \textrm{Hz}$, which in turn depends on the assumed breaking strain value of $\sigma \approx 0.05$ \citep{2025ApJ...978L...8M}. 
Crusts with a bigger breaking strain (up to $\sigma=0.1$ is possible in neutron stars \citet{2009PhRvL.102s1102H}) break at higher frequencies, which in our framework translates into fewer Sco X-1 compatible configurations that lead to crustal breakage, therefore negatively impacting the detection probability.

\end{description}

\subsection{Prospects}
\label{sec:cav_and_crit}

The framework introduced in this work can be used in future studies focusing on different aspects of Sco X-1. Enhancements to this framework could further refine our understanding of the prospects of detecting continuous gravitational waves from Sco X-1, help to better set up searches, and, one day, interpret observations. 
We outline below opportunities for future developments.

\begin{itemize}

\item Our binary evolutions all begin with a newly born neutron star and a companion that has just entered the main sequence. This is the same approach as taken by \citet{2021ApJ...922..174V} as well as \citep{2024MNRAS.535..344K, 2025ApJ...980...51K}, and it is perfectly consistent with the scenario of Sco X-1 originating from a pre-existing binary system where the initially more massive star goes through supernova explosion, the binary survives this event, evolving subsequently.
It cannot however be excluded that Sco X-1 might be the result of a close encounter (dynamical capture) with a single star or another binary system in a dense environment like a globular cluster \citep{2003AA...398L..25M}.
If this is the case, the captured companion star may find itself in any evolutionary stage, and the modelling of "Sco X-1 progenitor" systems becomes considerably more complex.

\item  When the spin evolution is dominated by the accretion torque, our simulations cannot resolve changes happening on timescales between  $10^4 - 10^5 \, \textrm{yr}$. This means that throughout the Sco X-1 compatible phase, we cannot resolve transitions between rapid spin-up and torque-balance. This effectively overestimates the time during which each simulated signal is emitting a detectable signal, ultimately affecting our detection probability estimates. This effect is likely very small for Sco X-1 configurations emitting at current detectable levels, as the vast majority of these are in torque balance or close to it.
Sco X-1 configurations with weak deformations emitting at $3G$ detectable levels might instead be affected. 
However, only for a narrow range of ellipticities, the switch between spin-up and torque balance happens during the Sco X-1 compatible phase; therefore, the impact of this effect will likely be negligible also in this case.

\item Spin-induced elastic deformations could be sourcing continuous gravitational waves from an accreting neutron star.
\citet{2024PhRvD.110d4016M} provide a model for this mechanism, with the resulting ellipticity depending on the degree of crustal anisotropy. The range of this quantity is not known. We have performed a preliminary investigation using the framework described in this paper, finding that the degree of crustal anisotropy necessary for a detection at current sensitivity must be higher than $1\%$. We intend to explore this, and other promising mechanisms relevant to accreting neutron stars (e.g., thermal mountains, \citet{2025ApJ...980..144L, 2025arXiv251010443L}), in greater depth in the future.

\item The size of the total possible Sco X-1 parameter space affects the estimates of the detection probabilities. Whereas orbital period and mass of the donor star are observationally constrained, the range of possible neutron star deformations is not, and is based solely on estimates.
Among the many assumptions made here, the range of ellipticities is what impacts most the detection probabilities. We have adopted a minimum value of $10^{-9}$ based on \citep{2018ApJ...863L..40W}, but the smallest ellipticity could well be a few/several orders of magnitude smaller \citep{Mastrano2011}, leading to detection probabilities correspondingly smaller. For example, if the smallest ellipticity happens to be $\varepsilon_{\textrm{min}} \approx 10^{-12}$, which is sometimes used as a benchmark for the lowest deformation in magnetically deformed neutron stars \citep{2008MNRAS.385..531H}, the parameter space available to Sco X-1 will be $\approx 1.7$ times bigger, and detection probabilities will decrease roughly of the same factor given that ellipticities smaller than $10^{-9}$ are essentially beyond detectability. 
We flag this as a point of caution in using our detection probabilities.

\item We did not model the propeller effect, which could impact our results if, at some point in its past evolution, Sco X-1 did undergo a propeller phase. If that happened, we would likely observe a transitional X-ray source, whereas Sco X-1 has been persistently accreting since it was first observed, suggesting that a propeller phase, if at all, might be in Sco X-1's future. Furthermore, an appropriate modelling of the propeller effect will substantially increase the dimensionality of the parameter space, and might dominate probability estimates just due to sheer ignorance of its mechanisms.

On the other hand, it is worth reflecting on the more realistic scenario of a time-varying accretion efficiency. At the onset of mass accretion we expect $\eta \approx 100\%$, with $\eta$ later on decreasing due to the many non-conservative mass transfer processes that can take place in Sco X-1, as mentioned in Section \ref{sec:mesa_grid}. It might be that the set of $P-M^{d}$ configurations evolving into Sco X-1 in this more complex scenario, would not be too far off from those found in the simpler case of constant $\eta$, as long as the time-averaged accretion efficiency is not $\lesssim 30\%$. 
We say this because we find that the sets of progenitors obtained at different fixed accretion efficiencies between $40\% \leq \eta \leq 80\%$, are roughly similar to each other.
Time-dependent accretion efficiency might instead significantly affect which Sco X-1 progenitors lead to the failure of the neutron star's crust. If indeed the $\eta \approx 100\%$ phase lasts long enough, crustal failure may also occur for those progenitor systems for which we find the crust does not break in the simplified case of constant $\eta$.

\end{itemize}

\section*{Acknowledgements}

MAP acknowledges Badri Krishnan and Yuanhao Zhang for earlier collaborative work exploring related ideas on the evolution of Sco X-1, which helped inspire the present, independent investigation. 
GP thanks N. Ivanova for greatly helping in providing the implementation of the CARB prescription in MESA.
GP thanks J.J. Ho Zhang for fruitful discussion about the accretion torque and its effects.
GP thanks A. Vargas for helpful discussion about the frequency time derivative tolerance of Sco X-1 searches.
DM acknowledges that this project has received funding from the European Research Council (ERC) under the European Union’s Horizon 2020 research and innovation programme (grant agreement No. 101002352, PI: M. Linares).

%%%%%%%%%%%%%%%%%%%%%%%%%%%%%%%%%%%%%%%%%%%%%%%%%%
\section*{Data Availability}

MESA inlists and data required to calculate detection probabilities are available on the official website of the Continuous Gravitational Waves group at the MPI/AEI Hannover \citep{supplementary_material}, as well as on the journal's webpage.
MESA inlists will also be available on ZENODO.



\bibliographystyle{mnras}
\bibliography{scox1_biblio} % if your bibtex file is called example.bib

%\begin{thebibliography}{99}
%\end{thebibliography}

%%%%%%%%%%%%%%%%%%%%%%%%%%%%%%%%%%%%%%%%%%%%%%%%%%

%%%%%%%%%%%%%%%%% APPENDICES %%%%%%%%%%%%%%%%%%%%%

\appendix

\section{Extra figures}
\label{sec:appendix_figs}

\begin{landscape}
	\begin{figure}
	\includegraphics[width=1.0\linewidth]{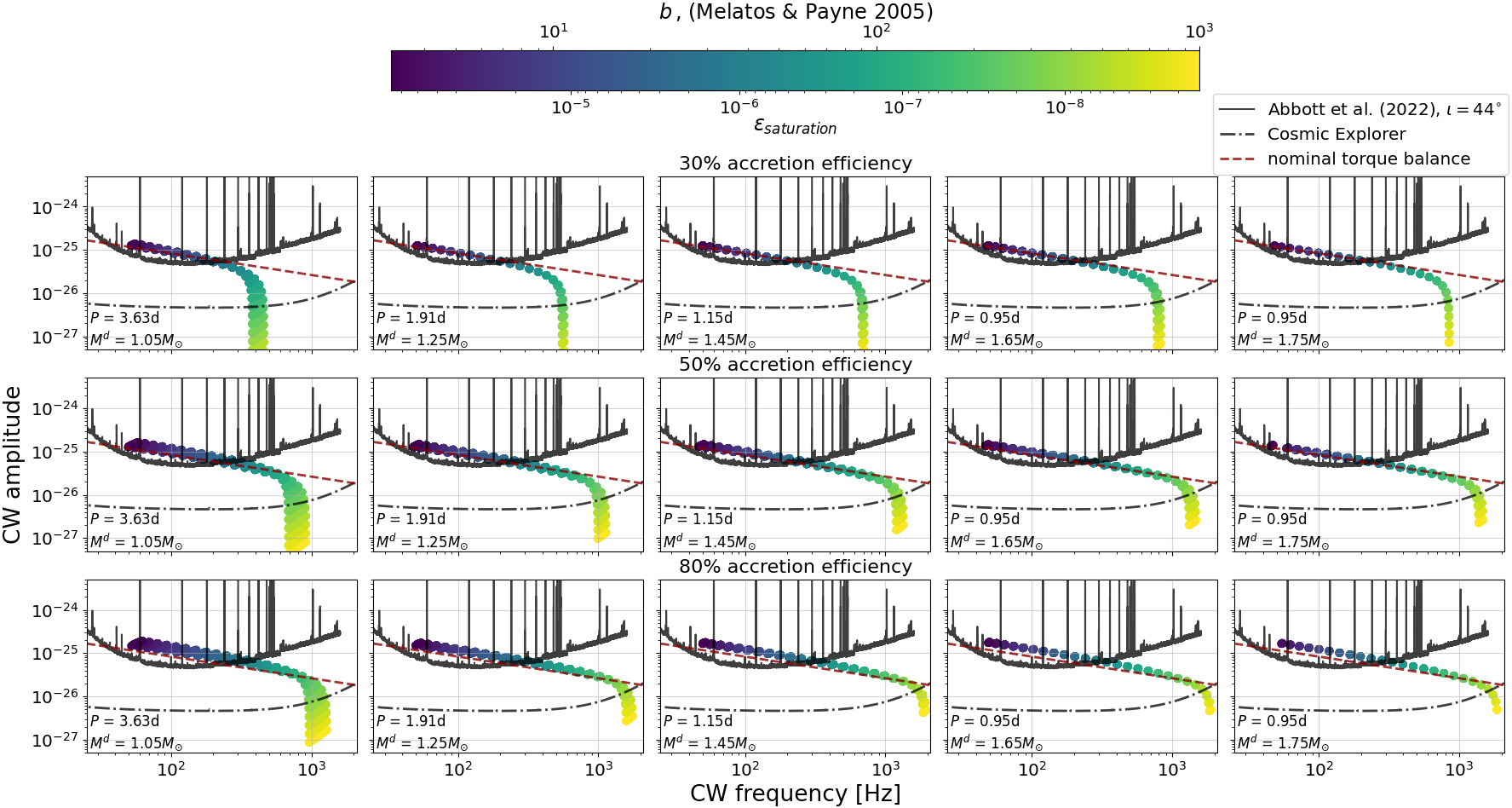}
	\caption{Continuous gravitational wave amplitudes $h_0$ against their frequency (purple to yellow traces) stemming from compatible Sco X-1 systems in the case of magnetic confinement of accreted matter for different progenitors (indicated in the bottom left corner of each panel), accretion efficiencies, and ellipticities. To facilitate visualisation, we only display gravitational waves emitted at evolutionary stages where the system resembles Sco X-1.
The chosen subset of binary parameters displayed is sufficient to appreciate the overall effect the various quantities have on the outcome of the simulations.Traces are colour-coded according to the value of $b$, or equivalently, to the saturation ellipticity.
	Amplitude upper limits (dark grey curve) are taken from \citet{2022ApJ...941L..30A}, assuming the spin axis of the neutron star to be orthogonal to Sco X-1's orbital plane.
	The dash-dotted line is the $3G$ projected sensitivity of \citet{2022ApJ...941L..30A} (Equation \ref{eq:h0_future} using $\mathcal{D} = 62$ and $N=1$).
	The red dashed line is the nominal torque balance curve.
	}
	\label{fig:h0_vs_f_mag}
	\end{figure}
\end{landscape}

\begin{figure*}
	\includegraphics[width=1.0\linewidth]{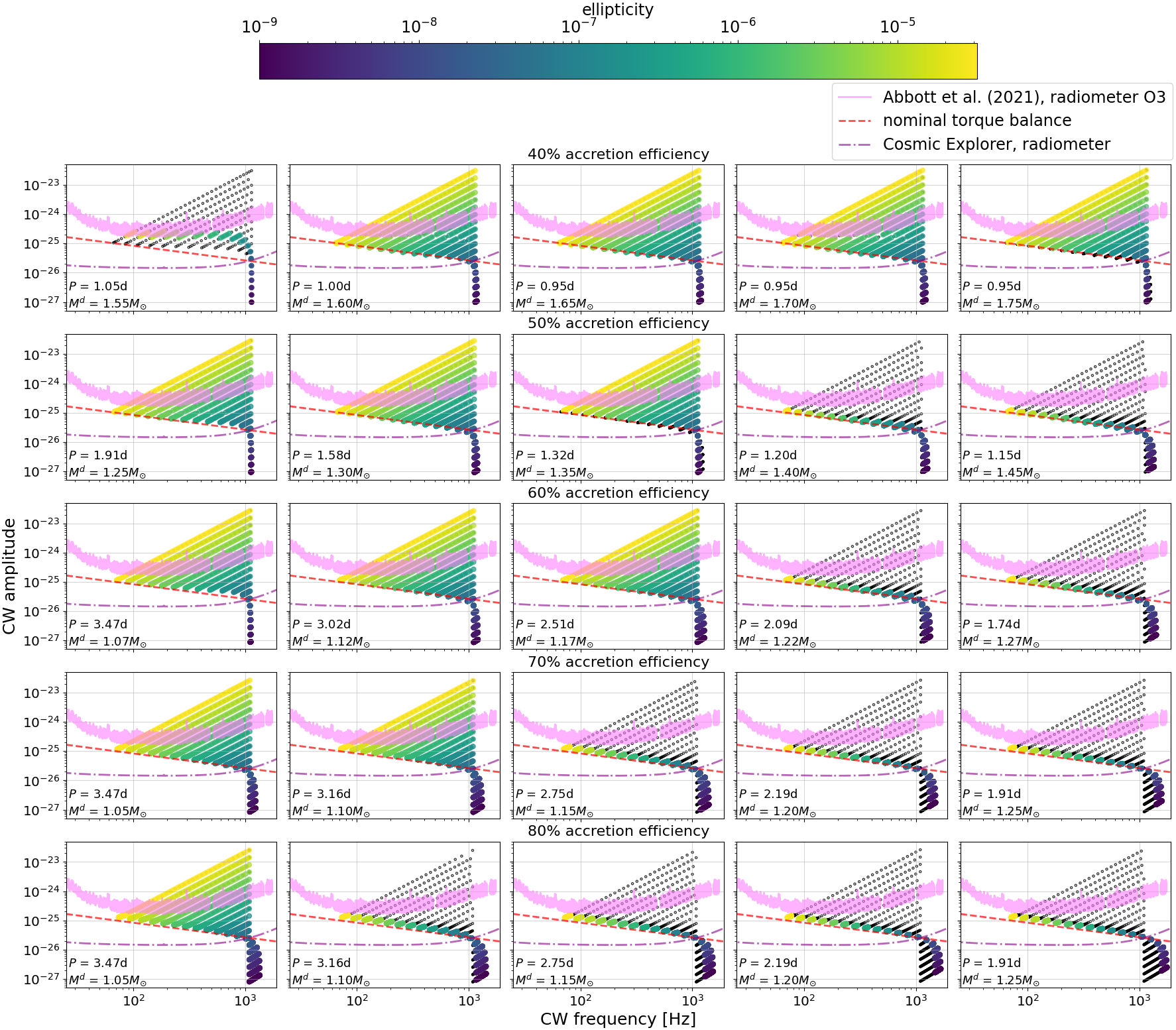}
	\caption{Continuous gravitational wave amplitudes $h_0$ against their frequency stemming from Sco X-1 systems in the case of residual ellipticity after crustal breakage for different progenitors (indicated in the bottom left corner of each panel), accretion efficiencies, and ellipticities (colour-coding of the filled scatter markers). Empty black circles are gravitational waves emitted while the Sco X-1 compatibility condition is not met. Filled scatter markers are gravitational waves emitted at evolutionary stages where the system resembles Sco X-1. Amplitude upper limits (magenta solid curve) are taken from LIGO O3 radiometer search \citet{2021PhRvD.104b2005A}.
	The dash-dotted line is the $3G$ projected sensitivity of the radiometer search (Equation \ref{eq:h0_future} using $\mathcal{D} = 14$ and $N=1$). The red dashed line is the nominal torque balance curve. At each accretion efficiency, the first panel from the left always refers to the Sco X-1 progenitor with the least massive donor leading to gravitational wave emission during the Sco X-1 compatible phase.}
	\label{fig:h0_vs_f_break}
\end{figure*}

%%%%%%%%%%%%%%%%%%%%%%%%%%%%%%%%%%%%%%%%%%%%%%%%%%

% Don't change these lines
\bsp	% typesetting comment
\label{lastpage}
\end{document}